# Review of fluid flow and convective heat transfer within rotating disk cavities with impinging jet

S. Harmand[1,2*], J. Pellé[1,2], S. Poncet[3], I.V. Shevchuk[4]

1 Université de Lille Nord de France, F-59000 Lille, France
2 UVHC, TEMPO-DF2T, F-59313 Valenciennes, France
3 Aix-Marseille Université, M2P2, UMR 7340 CNRS, F-13451 Marseille, France
4 MBtech Group GmbH & Co. KGaA, D-70736 Fellbach-Schmiden, Germany
* Corresponding author : souad.harmand@univ-valenciennes.fr

## Abstract

Fluid flow and convective heat transfer in rotor-stator configurations, which are of great importance in different engineering applications, are treated in details in this review. The review focuses on convective heat transfer in predominantly outward air flow in the rotor-stator geometries with and without impinging jets and incorporates two main parts, namely, experimental/theoretical methodologies and geometries/results. Experimental methodologies include naphthalene sublimation techniques, steady-state (thin layer) and transient (thermochromic liquid crystals) thermal measurements, thermocouples and infra-red cameras, hot-wire anemometry, laser Doppler and particle image velocimetry, laser plane and smoke generator. Theoretical approaches incorporate modern CFD computational tools (DNS, LES, RANS etc). Geometries and results part being mentioned starting from simple to complex elucidates cases of a free rotating disk, a single disk in the crossflow, single jets impinging onto stationary and rotating disk, rotor-stator systems without and with impinging single jets, as well as multiple jets. Conclusions to the review outline perspectives of the further extension of the investigations of different kinds of the rotor-stator systems and their applications in engineering practice.

# Nomenclature

| | | |
|---|---|---|
| Cd | Diffusion coefficient | $m^2.s^{-1}$ |
| Cp | Thermal capacity | $J.kg^{-1}.K^{-1}$ |
| Cw | Mass flow rate coefficient | |
| D | Jet diameter | m |
| e | Distance from rotor/ impinged surface to nozzle | m |
| f | function | |
| G | Dimensionless spacing interval | |
| h | Convective heat transfer coefficient | $W.m^{-2}.K^{-1}$ |
| $\bar{h}$ | Mean convective heat transfer coefficient | $W.m^{-2}.K^{-1}$ |
| K | Entrainment coefficient | |
| k | Mass transfer coefficient | $m.s^{-1}$ |
| $\dot{M}$ | Sublimation rate | $kg.m^{-2}.s^{-1}$ |
| L | Reference length | m |
| Q | Volumetric flow rate | $m^3.s^{-1}$ |
| q | Heat Flux | $W.m^{-2}$ |
| n | Exponent in the power-law approximation of the temperature profiles | |
| r | Local radius | m |
| R | External radius | m |
| t | Time | s |
| T | Temperature | K |
| u, v, w | Time-averaged velocities | $m.s^{-1}$ |
| u', v', w' | Turbulent velocities | $m.s^{-1}$ |
| U | Cross-flow velocity | $m.s^{-1}$ |
| V | Reference Velocity | $m.s^{-1}$ |
| W | Jet Velocity | $m.s^{-1}$ |
| z | Axial direction | m |

*Greek Symbols*

| | | |
|---|---|---|
| $\alpha$ | Thermal diffusivity | m²/s |
| $\beta$ | core-swirl ratio | - |
| $\Delta$ | Variation | |
| $\delta$ | Boundary layer thickness | m |
| $\kappa$ | Flow parameter | |
| $\lambda$ | Thermal conductivity | $W.m^{-1}.K^{-1}$ |
| $\rho$ | Density | $kg.m^{-3}$ |
| $\Omega$ | Rotational velocity | $s^{-1}$ |
| $\tau$ | Wall shear stress | $N.m^{-2}$ |
| $\nu$ | Kinematic viscosity | $m^2.s^{-1}$ |

*Upper-Subscripts*

| | |
|---|---|
| a | Air |
| ad | Adiabatic |
| lam | laminar |
| r | Of reference |
| s | Naphtalene Solid |
| tur | Turbulent |
| v | Naphtalene Vapor |
| w | Wall |
| r,θ,z | Radial, Tangential, Axial |

*Dimensionless Numbers*

| | |
|---|---|
| $\theta$ | Dimensionless temperature |
| $C_M$ | Moment coefficient |
| Fo | Fourier number |
| G | Dimensionless airgap spacing interval |
| Gc | Dimensionless shroud clearance |
| Gr | Grashof number |
| $Nu_D$ | Local Nusselt number based on jet diameter D |
| $Nu_r$ | Local Nusselt number based on local radius r |
| $\overline{Nu_R}$ | Mean Nusselt number based on external radius R |
| Ra | Rayleigh number |
| $Re_e$ | Reynolds number, based on $\Omega$ and e. |
| $Re_\Omega$ | Rotational Reynolds number, based on $\Omega$ and R. |
| $Re_r$ | Local rotational Reynolds number based on $\Omega$ and r. |
| $Re_j$ | Jet Reynolds number based on W and D. |
| $Re_U$ | Cross-flow Reynolds number |
| Ro | Rossby number |
| Pr | Prandtl number |
| Sc | Schmidt number |
| Sh | Sherwood number |
| $u^*, v^*, w^*$ | Dimensionless velocities |
| $z^*$ | Dimensionless axial direction |

# Introduction

Fluid flow and convective heat transfer in rotor-stator configuration are of great importance in engineering practice in the turbomachinery industry, namely, in the gas turbine cooling. Another important example from the modern practice has relevance to wind generators. Discoidal rotor-stator systems do not incorporate gears in order to avoid altering the system's mechanical efficiency, and are thus able to improve power production at low rotational speeds. Reliable methods for estimation convective heat transfer intensity in the air gap can help manufacturers to better design the discoidal wind generators to prevent overheating. Different strategies are employed by designer depending on the particular engineering application.

A review of the most important results for the rotor-stator systems available so far is presented here. The review focuses on convective heat transfer in predominantly outward air flow in rotor-stator cavities with and without impinging jets. The structure of the review comprises two main parts, namely, methodology and geometries/results.

Methodologies elucidate experimental techniques most widely used for measurements in rotating-disk systems of (a) heat/mass transfer, such as naphthalene sublimation techniques, steady-state (thin layer) and transient (thermochromic liquid crystals) thermal measurements, thermocouples and infra-red cameras, and (b) flow structure using hot-wire anemometry, laser Doppler and particle image velocimetry, as well as by means of a laser plane and smoke generator. Numerical methodologies for flow and heat transfer modelling in the rotor-stator systems encompass methods of the direct numerical simulation (DNS), Reynolds-averaged Navier-Stokes equations (RANS) with the turbulence models both in the general-purpose form as incorporated in the commercial computational fluid dynamics (CFD) software, and often specially modified to be used in the rotating-disk systems. They include different modifications of standard $k$-$\varepsilon$, $k$-$\omega$ and Reynolds-stress models (RSM); large-eddy simulations (LES) approach is also quite widely used.

Geometries and results part is built up on the principle of ascent from simple to complex geometries. This helps understand complex fluid flow and heat transfer phenomena in the rotor-stator systems as a superposition of a number of separate effects like rotation, impingement/crossflow with different angles of attack, flow regime, wall temperature profile etc.

It starts with a free rotating disk. Both fundamental findings and recent advancements in the theoretical and experimental investigations of fluid mechanics and heat transfer over a free rotating disk in laminar, transitional and turbulent flow regimes are presented here together with the recommendations on the most reliable relations for prediction of the friction and transport coefficients including effect of the radial profile of the disk temperature.

The next more complicated geometry is the single disk in the cross-flow. Here experimental/computational results are discussed for a stationary/rotating disk of a finite and zero thickness in parallel cross-flow including the case where the disk was mounted on a side surface of cylinder with a significantly smaller radius.

Single jets impinging onto stationary and rotating disks are dealt with in the next part in both experimental and theoretical aspects involving self-similar exact solution for the stagnation point.

The following section describes in details peculiarities of fluid flow and heat transfer in the rotor/stator system without impinging jets. Presented here are maps of the flow regimes depending on the height of the gap between the rotor and stator and rotational speed. The section is completed with equations for the moment coefficient and Nusselt number.

As the next step, rotor-stator systems with the impinging jets are treated. Flow regimes including Batchelor and Stewartson flow structures are elucidated here, and predictive relations for the boundaries of the flow regimes, friction coefficients and Nusselt numbers complement this material.

In the end, rotor-stator systems with the multiple impinging jets finalize the description of the geometrical configurations in the review.

Conclusions to the particular parts of the review outline perspectives of the further extension of the investigations of different kind of the rotor-stator systems and their and their applications in engineering practice.

# 1. Methodology

## 1.1. Experimental methodologies for flow and heat transfer

In this part are considered methodologies which are used in studies about rotating disks with or without jet impingement, impinging jet on stationary surfaces, rotor-stator configurations with or without jet.

### 1.1.1. Convective Heat Transfer

**Mass transfer.** The general naphthalene sublimation technique for mass transfer measurements is described in detail in [SOU91] and [GOL95]. The surface where heat and mass transfer is to be determined is covered with a thin solid layer of naphthalene. Different methods are available for coatings, but the casting procedure is the most convenient and popular method [GOL95]. The coating thickness is usually less than 0.2mm. The air flow near the surface involves a mass and heat transfer from the wall to the air. Some works [KRE68, SPA82, TIE63, JAN86, WAN11] use that technique to determine a surface-averaged mass transfer on an entire disk by only measuring the weight before and after experiments and determining the global sublimation rate. To do that, a 0.1 mg resolution balance is often used with a 200g capacity [GOL95]. In order to ensure accurate measurements, the run time must be such that at least 10mg of naphthalene sublimate. Other works [CHE98, CHO03, HE05, CHO01, MAB72, PRA95, SHI87] experimentally study the local convective heat transfer in rotating disk configurations using this technique. Then the mass transfer coefficient $k$ is determined with:

$$k = \frac{\dot{M}}{\rho_v - \rho_a} \tag{1}$$

where $\rho_v$ and $\rho_a$ are the naphthalene vapor and air density, and $\dot{M}$ is the sublimation rate. The mass transfer is essentially isotherm and the naphthalene vapor density is assumed to be constant and equal to the one corresponding to the cooled surface temperature. In order to determine $k$ precisely, the sublimation rate $\dot{M}$ has to be determined by evaluating the sublimation thickness $\Delta z$ due to the mass transfer:

$$\Delta z = \frac{\dot{M}\Delta t}{\rho_s} \tag{2}$$

where $\rho_s$ is the naphthalene solid density; $\Delta t$ is the duration of the sublimation; $\Delta z$ is the naphthalene thickness, which is sublimated during the experiment. The most accurate method for the thickness measurement is the use of a linear variation differential transformer (LVDT). An LVDT having a 0.5mm linear range and 25nm resolution is well-adapted. Different tips are used for the LVDT gauge. The tip is selected so that it can measure a slope change of the surface with little error. A tip pressure that is too high produces a scratch mark on the naphthalene surface, whereas a tip pressure that is too low can introduce more measurement error [GOL95].

The measured $\Delta z$ depends on the position over the disk. The local mass flux is then evaluated with Eq. (2) and $k$ is deduced with Eq. (1). The local Sherwood number $Sh$ is then determined by:

$$Sh = \frac{k.L}{Cd_v} = \frac{\Delta z.s.\rho_s}{(\rho_v - \rho_a).Cd_v.\Delta t} \tag{3}$$

with L a reference length, the air gap thickness in the case of a rotor stator system [PRA95] or jet diameter in [CHE98], and $Cd_v$ the diffusion coefficient of the naphthalene vapor in the air, which is usually obtained with an equation of the following form:

$$Cd_v = a.T^b \tag{4}$$

where T is the air temperature inside the gap or near the disk. Depending on the author, a ranges from $8.1771 \times 10^{-7}$ to $3.0933 \times 10^{-6}$ and b from 1.75 to 1.983. The Reynolds analogy is then used to link the mass transfer to the heat transfer using the local Nusselt, Sherwood, Schmidt and Prandtl numbers:

$$Nu = Sh\left(\frac{Pr}{Sc}\right)^m \tag{5}$$

The exponent m is independent of the Pr and Sc numbers, with the values of m being determined from the empirical results for different flow regimes. Typically, the Schmidt and Prandtl numbers in different works varied over the range Sc=[2.28-2.5] and Pr=[0.7-0.74], respectively.

An in-depth investigation into the proper values of the exponent m based on the experimental data and theoretical models of different authors was performed in [SHE08, SHE09a]. For laminar flow, the value of m=0.53 first suggested by [JAN86] was confirmed by the authors of [SHE08, SHE09a]; for transitional flow, m=0.6 in the narrow range of $Re_\Omega$=[1.9×10$^5$-2.75×10$^5$] [SHE08, SHE09a]; for turbulent flow, Eq. (5) was modified as $Nu/Sh = Pr^{0.64}/Sc^{0.96}$ [SHE08, SHE09a]. The value m=0.4 used for all flow regimes in e.g. [CHE98, HE05, WAN11] in analogy with a non-rotating flat surface entails significant errors in recalculating the experimental Sherwood numbers to the respective Nusselt numbers using the analogy theory (Eq. (5)).

Sensibility of the exponent m to the flow regime and thus indirectly to the geometrical configuration that affects the flow regime makes the naphthalene sublimation technique rather inconvenient, when the task of determining heat transfer coefficients is predominant. Also, though it was rather widely used, this technique is submitted to limitations in the case of low velocities due to the need of increasing the time of experiments which induces inaccuracies. And in the case of high velocities, the problem is that naphthalene can be worn or stripped of by mechanical erosion rather than diffusion and convection.

Electrochemical techniques employing rotating-disk electrodes were used in [DOS76, MOH76, SAR08] for deduction of the mass transfer coefficient and then the Sherwood number with the ultimate objective to experimentally find out diffusion coefficients of particular electrolytes. As discussed in detail in [SHE09a, SHE09b], for the high Schmidt numbers specific for electrochemistry applications, diffusion boundary layer is extremely thin and develops within the viscous sub-layer of the velocity boundary layer. This destroys full analogy between heat and mass transfer observed in the case of the naphthalene sublimation technique, entails e.g. higher values of the exponent at the Reynolds number for turbulent flow (0.9 instead of 0.8 for heat transfer applications) and makes it hardly possible to use electrochemical techniques to model heat transfer based on the analogy (Eq. (5)).

**Thermal measurements.** At present, most of authors use direct thermal measurements to determine the convective heat transfer. With the purpose to determine the local or mean heat transfer coefficient between radii $r_1$ and $r_2$ over a disk, it is usually defined as:

$$h = \frac{q}{T_w - T_{ad}} \tag{6}$$

$$\bar{h} = \frac{2\int_{r_1}^{r_2} q \cdot r \cdot dr}{(r_2^2 - r_1^2)(\overline{T_w - T_{ad}})} \tag{7}$$

With:

$$\overline{T_w - T_{ad}} = \frac{2\int_{r_1}^{r_2} (T_w - T_{ad}) \cdot r \cdot dr}{(r_2^2 - r_1^2)} \tag{8}$$

The local convective heat flux q can be obtained in several ways, depending on the author. Most of them [LYT94, CAR96, SAN97, LEE02, GAO06, LIU08, KAT08, AST08] performed *steady-state experiments* and used a thin layer technique by depositing a thin heat foil on the cooled surface, which delivers a known and uniform Joule heat flux during experiments. Thin layer technique enables to assume that a 1-D resolution (i.e. heat conduction model in the direction orthogonal to the measurement surface) is sufficient, neglecting lateral conduction inside disk. To our knowledge, the use of this technique associated with a multidimensional resolution was only made by [LIU08]. The opposite face of the heat foil is usually insulated in order to decrease heat losses by the backside of the disk, except for [LYT94] where convective losses occurred and were estimated and taken into account with available correlations. The Joule heat flux was also corrected via taking into consideration an estimated radiative heat flux on the cooled surface in [LYT94, CAR96, LEE02, GAO06] or for both faces in [KAT08]. No corrections were used in [SAN97, GUE05], however related errors involved due to this were estimated to be under 5%. Owen [OWE74] and Pellé & Harmand [PEL07, PEL09, PEL11] worked with the thick layer technique. In those experiments, the disk was heated on the backside by infrared emitters in [PEL07, PEL09, PEL11] or by thermal resistances [OWE74] and cooled by forced convection on the front face where measurements were performed. In the method used by Pellé & Harmand [PEL07], the cooling surface of a rotating aluminum disk was covered with an insulating material. Local convective heat transfer variations lead to radial temperature variations over the cooled surface. These variations in amplitude depend on the nature of the flow, the thermal conductivity of the material, and the thickness of the insulating layer. The conductive heat flux was calculated by solving the heat conduction equation by a finite difference method using measured temperatures as boundary conditions, after reaching a stationary state. Radiative heat flux was then estimated and subtracted to compute the pure convective heat flux in [PEL07], while [OWE74] assumed the radiative heat flux to be negligible.

*Transient techniques* were also used in [YAN97, SANI98, AXC01, KAK09] using an analytical solution of the one-dimensional conduction with known initial and boundary conditions in order to restore the local heat transfer coefficient. In the works of [YAN97, SANI98], the disk was heated in an oven before experiments until reaching the stationary thermal state, while in [AXC01] or [KAK09] the surrounding fluid is heated. Via recording of the wall temperature cooling/heating history in a short time period (usually up to 1 min) the heat flux was restored using the aforementioned one-dimensional heat conduction solution.

The measurement of the cooled/heated surface temperature $T_w$ can be achieved in 3 main ways.

Few studies used the *Thermochromic Liquid Crystals* (TLC) [YAN97, SANI98, AXC01, LEE02, KAK09], namely in transient techniques. The main limitation of this technique is the precise calibration that it is required. The first obvious restriction is that heat conduction must involve only a small fraction of the real wall thickness for the 1D heat conduction solution for a semi-infinite-slab to hold. According to [SCH73, IND04, SHE06], the Fourier number $Fo = \frac{4\lambda t}{L^2}$ must not exceed the values 0.25 up to 0.3, where $\alpha=\lambda/(\rho Cp)$ is thermal diffusivity of the wall, *L* is the thickness of the wall, *t* is the measurement time. For instance, for a disk of a thickness *L*=0.01 m made of Plexiglas® [YAN97, SANI98, IND04, SHE06], the measurement time in transient experiments must not exceed 69 sec. The second obvious restriction is that the shape of the initial temperature non-uniformity over the disk surface must hold during the time of measurements. According to [SHE06, SHE09a], this is fully fulfilled for a disk made of Plexiglas® in view of the applicability of the semi-infinite-slab approach for a wide range of both radially increasing and decreasing temperature distributions over the disk surface (including the isothermal disk). For a disk made of highly conductive material, like e.g. Aluminum, the transient experimental technique is inapplicable for all kinds of the initial surface temperature non-uniformity, except for the case of an isothermal disk [SHE06, SHE09a].

*Thermocouples* were used in [OWE74, SAN97], where the disk was made of a relatively high thermal conductivity material, in order to assume that the measured temperature inside the disk was the same as the surface temperature. The thermocouples must be also placed with a thermally-well conductive paste in order to avoid contact resistance. With the use of thermocouples on rotating systems, it is very difficult to get a sufficient spatial resolution. A rotating mercury-ring collector must be placed to transmit the signal from the rotating disk to the stationary acquisition system, with the number of channels being limited. Techniques involving the thermocouples and partially the TLC are intrusive, because they can affect the thermal properties of the cooled/heated body.

That is why most of authors use *infrared thermography* [LYT94, CAR96, BOU05, GAO06, PEL07, AST08, KAT08, PEL09, PEL11]. Usually, the cooled surface is covered with high emissivity black paint ($\varepsilon \geq 0.95$). The main difficulty of this technique is to clearly identify the infrared signal emitted from the observed surface, namely in the case of a confined configuration such as rotor-stator configurations, where reflections can occur. Moreover, in confined configurations, a measurement window with high transmittive materials (zircon or germanium) that requires careful calibration must be used as pointed out in [KAK09]. A detailed description of the relevant methodology is documented in [PEL07, PEL09].

The adiabatic wall temperature $T_{ad}$ is the temperature that would be the surface temperature without any heat transfer, which would be also the fluid temperature near the wall. It takes into account possible overheating due to viscous friction at the wall which depends on the fluid dynamics [KAK09]. Usually, it is written as:

$$T_{ad} = T_r + \frac{f(V^2)}{2.Cp} \qquad (9)$$

with $T_r$ being the temperature of the fluid entrained by the surface boundary layer and $V$ being a characteristic velocity representative for the considered flow, and f being an increasing function with V. For a laminar flow on a rotating disk, the adiabatic temperature can be easily calculated using the theory developed in the book by Owen [OWE89], so that $f(V^2)=Pr^{1/3}(\omega r)^2$. Cardone et al. [CAR97] have measured the adiabatic wall temperature for high rotating speed up to 21000 rpm. For those velocities, their results are in agreement with the theory by Owen [OWE89] and exhibit a temperature rise up to 10K only due to viscous dissipation. A method for determination of $T_{ad}$ was also proposed by Newton [NEW03]. In most papers, tangential velocities ($\omega r$) are not so high than in [CAR97], so that the authors consider that the error introduced by considering $T_{ad} = T_r$ is negligible in comparison with the difficulty of its measurement. So, $T_r$ is often chosen as the characteristic (reference) temperature of the fluid in which the experiment is made: ambient (or jet) air temperature in the case where the measurement surface is heated [LYT94, YAN97, SANI98, LEE02, PEL07, LIU08, PEL09, PEL11] or fluid temperature when the fluid is heated in place of the disk [AXC01]. According to [CAR97], neglecting the influence of the viscous dissipation can lead to some errors in the determination of the Nusselt numbers, namely at outer radii. The choice of the reference temperature is a very difficult point, and in order to compare the different studies of heat transfer the reference temperature measurement must be presented and justified.

### 1.1.2. Flows

**Hot wire anemometry.** Hot wire anemometry is a well-known method used to measure velocity profiles and Reynolds stresses in a complex flow. However, it is an intrusive method used by several authors in rotating [OWE74, ITO98, DJA98, DJA99, DEB10] or jet impingement [ITO98, GUE05] configurations.

In [DJA98, DJA99], radial and azimuthal velocity components u and v as well as the turbulent correlations $u'^2$, $v'^2$, $u'v'$ were measured. The technique involved using a special probe made of three wires, 2.5 µm in diameter, all situated on the same plane and having an angular spacing equal to 120°. In [DEB10], the authors used the same technique to measure u, v and $v'^2$, in an opened rotor-stator cavity with an inward radial flow.

In [ITO98], two hot wires were used: one had a wire mounted normal to its axis of rotation and capable of measuring the components u, v and $u'^2$, $v'^2$, $u'v'$; the other had a wire inclined at an angle of 50° to its rotation axis and was used for measurements of $w'^2$, $u'w'$ and $v'w'$.

The determination of the mean flow field in the case of an impinging jet was made with the use of 10000 samples in [GUE05], and each velocity profile was established with 100 measurement points. Authors also indicated that the measurement becomes independent of the number of samples for more than 5000 samples.

**Laser Doppler Velocimetry.** In the Laser Doppler Velocimetry (LDV) or Laser Doppler Anemometry (LDA), the difference between the frequencies of the initially sent laser signal and received signal after passing through the flow containing particles is estimated, which enables deducing velocity values and directions? Used particles are glycerine (diameter of about 5µm) [DON04], olive oil (diameter of about 1µm) [KOS07], salt-water solution (diameter of about 1µm) [LYT94] or corn statch (diameter of about 5 µm) [FIT97].

Poncet et al. [PON05a, PON05b] used this experimental technique in a rotor-stator configuration with throughflow. Particle type was "Optimage PIV seeding powder" with the particle size of about 30µm. As pointed out, the main defect of this method is to provide an integrated value over the probe volume, whose size in the axial direction can be large compared to the air gap interval or at least to the boundary layer thicknesses. The rotor-stator cavity in which Geis [GEI01] made flow measurements is double-shrouded with a perturbed rotor surface. [LYT94, FIT97, KOS07] used this technique to study effects of a jet impingement flow on a flat stationary surface. In [DON04], for an impinging air jet configuration, areas of great interest were studied with LDV/LDA, with most of the entire flow pattern being investigated with the help of Particle Image Velocimetry.

**Particle Image Velocimetry.** Particle Image Velocimetry (PIV) is based on the displacement measurement of particles during two laser flashes, whose time between pulses is known. However, it usually provides less spatial and temporal resolution than LDA/LDV techniques, except when using high speed laser and cameras. [DON04] used PIV to have an idea of the main flow field before using LDA/LDV to closely look at the jet impingement area. In these experiments, the filed size was 1280×1024 pixels and minimum time between the frames was 200ns.

Boutarfa [BOU05] measured the mean flow field in different planes between a rotor facing a stator using a water spray added to the air flow. The dimensions of the field covered by the objective were 114×110 mm$^2$ and the thickness of the laser plane was about 5mm. The cross-correlation between two images was carried out on a 850×150 pixels zone. The time interval Δt between images, regulated as a function of the air velocity, ranged between 30 and 400 ns.

On a co-rotating disk configuration, Wu [WU09] used PIV with polycrystalline particles (diameter of about 30µm) with a 1.5mm thickness laser plane and a maximal acquisition frequency of 30Hz.

In LDV/LDA or PIV, the authors interested in flows in confined configurations (such as rotor-stator etc.) [BOU05, WU09] needed to use Plexiglas windows for optical access, whose thickness is usually under 10mm.

**Basic Visualizations.** Basic flow visualizations by means of a laser plane and smoke generator were used in [SOO02, HSI06] for configurations with two disks rotating independently and jet impingement onto a confined disk. Mean flow structure was identified and linked to heat transfer in [SOO02]. The transient experiment performed in [HSI06] enabled to highlight the time-dependent state of the flow and boundaries separating various flow regimes between the disks.

### 1.2. Numerical methodologies for flow and heat transfer modelling

Despite their relative geometric simplicity, enclosed rotating disk flows contain a complex physics, which makes their modelling a very challenging task for numerical methods. The flow problem presents indeed several complexities, such as the high rotation rates, confinement effects, the coexistence of laminar, transitional and highly turbulent flow regions, three-dimensional precessing vortical structures, very thin 3D turbulent boundary layers along the disks, strong curvature of the streamlines, recirculation zones etc. When an axisymmetric jet is superimposed on this flow and impinges onto the rotating disk in a confined rotor-stator system, the flow pattern becomes even more complex with the interaction between the jet and the secondary rotor-stator flow. Numerically, there are also some additional technical constraints due to firstly the singularity in the Navier-Stokes equations when dealing with cylindrical coordinates in cylindrical (and not annular) geometries and secondly the choice of the boundary conditions at the outlet to ensure the mass conservation. Thus, most of the studies up to now have been dedicated to either enclosed rotor-stator flows without jet or impinging jet flows on a flat plate without rotation and more rarely to both problems at the same time.

### 1.2.1. Direct numerical simulations (DNS)

To the best of our knowledge, three-dimensional DNS are still used to investigate the transition to turbulence in idealized enclosed cavities without throughflow under isothermal conditions [SER01, SER02, RAND06, PON09b]. Serre *et al.* [SER04a] extended their former work to investigate the effects of thermal convection under the Boussinesq approximation for Rayleigh numbers up to $2\times10^6$ for slightly different flow parameters (G=e/R=0.256, $Re_\Omega=1.1\times10^5$). These DNS were based on the same pseudo-spectral method with Chebychev polynomials in the non homogeneous directions and Fourier series in the tangential direction, which provides very highly accurate results but in relatively simple enclosed annular geometries. Moreover, they require several millions of mesh points and excessive calculation times preventing from performing a parametric study in such a configuration. We can cite also Lygren and Andersson [LYG01], who performed a DNS of the turbulent flow ($Re_\Omega=4\times10^5$) in an angular section of an unshrouded rotor-stator cavity assuming a periodicity of the flow in the tangential and radial directions.

### 1.2.2. Axisymmetric Reynolds Averaged Navier-Stokes (RANS) computations

In the turbulent regime, most of the numerical studies have assumed the flow pattern as being axisymmetric and steady. During the 1980s, flow modelling was based on axisymmetric RANS computations based on the isotropic linear eddy-viscosity hypothesis. Chew [CHE85] employed the mixing-length hypothesis to model the flow over single rotating disks. Morse [MOR88] used the low-Reynolds-number $k-\varepsilon$ model of Launder and Sharma [LAU74] over the complete cavity, which significantly improved the results compared to mixing-length models. Morse [MOR91] replaced the local turbulence Reynolds number in the sublayer damping functions by the normalized wall distance to avoid the rotor boundary layer to display an excessive relaminarization region. At the same time, Iacovides and Theofanopoulos [IAC91] used an algebraic modelling of the Reynolds stress tensor in the fully developed turbulence area and a mixing-length hypothesis within the sublayers. This non-linear eddy viscosity model provided satisfactory results in the case of a rotor-stator flow with and without throughflow but some discrepancies remained and the authors concluded that no single form of their models was satisfactory for all rotating disk configurations considered.

The advent of second-moment closures started in the 1990's in such flow configurations mainly because the direct effects of rotation on the individual Reynolds stress tensor components may be directly accounted for. Elena and Schiestel [ELE93b] proposed a first version of second-moment closure applied to rotor-stator disk flows. The first second moment closure taking into account the realizability diagram of Lumley [LUM78] in the framework of rotating flows was proposed by Launder and Tselepidakis [LAU94]. The third version of the low-Reynolds number Reynolds Stress Model of [ES96] has shown to improve their earlier model [ELE93b], when comparing with previous measurements. This model is sensitized to some implicit effects of rotation on turbulence such as an inverse flux due to rotation, which impedes the energy cascade, inhomogeneous effects, the spectral jamming and an additional contribution to the pressure-strain correlation, introducing the dimensionality tensor. This model was very favourably compared for a wide range of flow conditions including flow regimes with merged [HAD08] and unmerged boundary layers, with and without superimposed centripetal or centrifugal throughflow [PON05b] and heat transfer effects [PON07]. If it provides a correct distribution of laminar and turbulent regions and improved significantly the predictions of other RANS models, the Reynolds stress behaviour is not fully satisfactory, particularly near the rotating disk and a slight tendency for relaminarization close to the axis can be observed in the closed cavity case.

Da Soghe *et al.* [DASO10] compared several two-equation models available within the CFD code CFX to the RSM model of Elena and Schiestel [ES96] and experimental data for the same flow conditions as in Poncet *et al.* [PON05b]. They tested in particular two innovative versions of the k-ω SST model. The first one was proposed by Smirnov and Menter [SMI08], which include the modification proposed by Spalart and Shur [SPA97] to sensitize the model to the rotation and curvature effects. In the second modified version, an additional source term is introduced in the turbulence kinetic energy transport equation to improve the prediction of the reattachment location in the case of a superimposed outward jet. For industrial applications, the k-ω SST model with the reattachment modification slightly improves the predictions of the standard model and seems to offer a good compromise between accuracy and calculation cost even if the RSM model of [ES96] still provides the best predictions.

The axisymmetric jet impinging on a flat plate without rotation has served for many years as a fundamental test case for the investigation of the performance of turbulence models. Craft *et al.* [CRA93] compared four turbulence models with available experimental data in the case of turbulent impinging jets discharged from a circular pipe. The k-ε eddy viscosity model and one of three Reynolds stress models lead to too large turbulence intensities close to the stagnation point and as a consequence to too high heat transfer coefficients. The two second-moment closures, accounting for the wall effect on pressure fluctuations, provided more satisfactory results even if none of the schemes appeared to be entirely successful in predicting the effects of jet Reynolds number. More recently, Zuckerman and Lior [ZUC06] discussed in details the suitability of different RANS models for the modelling of impinging jet flow and heat transfer without rotation. Two-equation models failed to predict the impinging jet transfer coefficient and also the secondary peak in Nusselt number. RANS models based on the eddy-viscosity concept were also unable to catch the occurrence of negative production of turbulence kinetic energy close to the wall in the stagnation region. The modelling of this term using explicit algebraic stress models or second-moment closures remains again a challenging task. The authors showed that the $\overline{v^2}$ – *f* model can produce better predictions of fluid properties in impinging jet flows and may be recommended as the best compromise between solution speed and accuracy. Jaramillo *et al.* [JAR08] studied also the suitability, in terms of accuracy and numerical performance, of eleven RANS models (two-equation linear and non-linear eddy-viscosity models, explicit algebraic Reynolds stress model within k-ε and k-ω platforms) in the description of plane and round impinging jets. The authors mourned the lack of generality of the models. Nevertheless, they pointed out that the use of better tuned damping functions and additional terms such as the Yap correction in the length-scale-determining equation seemed to play an important role. The k-ω models produced in these different configurations less scattered results than those obtained by the k-ε models.

Recently, the 13th ERCOFTAC/IAHR Workshop on Refined Turbulence Modelling has focused its attention to the turbulence modelling of a round jet impinging perpendicularly onto a rotating heated disk [STEI09] in an opened computational domain. RANS predictions were compared to the experimental data of Minagawa and Obi [MIN04] and Popiel and Boguslawski [POP86] for the hydrodynamic and thermal fields respectively. The main flow was assumed to be axisymmetric and steady when dealing with RANS models. It appeared crucial to simulate the flow inside the inlet nozzle either by multidomain approach or by precursor computations of the fully developed pipe flow to have realistic conditions at the nozzle exit. In this context, Manceau *et al.* [MAN09] compared two in-house, unstructured, finite-volume codes together with five turbulence models to evaluate the model performances and to investigate the effect of rotation on turbulence and heat transfer. Heat transfers were taken into account through a simple gradient hypothesis with a turbulent Prandtl number ranging between 0.91 and 1. For all five models, they noticed a poor prediction of the velocity field in the jet region and attributed it to an overestimation of the free-jet spreading before its impingement on the rotor. The two RSM models (an elliptic-blending RSM and the RSM using scalable wall functions) did not improve significantly the results of the k-ω SST, ɸ-*f* and ζ-*f* models, unless if we consider the radial distribution of the radial fluctuating velocity close to the disk. They concluded by the need for additional modelling development to sensitize RANS models to the effect of rotation, which has been done only by Elena and Schiestel [ES96] for a Reynolds Stress Model.

### 1.2.3. Three-dimensional computations in the turbulent regime

Though the geometry in this kind of problem is strictly axisymmetric, some numerical simulations [TUC02, CRA08] and experimental flow visualizations [CZA02] have revealed the existence of precessing large scale vortical structures even in the turbulent regime. Owen [OWE00] has drawn attention to difficulties in predicting the flow in rotating disk arrangements using bidimensional steady calculations and speculated that the cause might be related to the formation of these vortices. Thus, Craft *et al.* [CRA08] developed a 3D unsteady version of a low-Reynolds number k-ε model and applied it to a turbulent rotor-stator flow in an enclosed cavity. Their results confirmed the "S-shaped vortex" regime but this strategy required a very fine mesh across the sublayers and also prohibitive time integration preventing from further calculations.

In the 2000's, Large Eddy Simulations (LES) appeared to be the appropriate level of modelling in such complex flows including 3D effects. Wu and Squires [WU00] performed the first LES of the turbulent boundary layer flow over a free rotating disk in an otherwise quiescent fluid at $Re_\Omega = 6.5 \times 10^5$. They concluded that when the grid is fine enough to accurately resolve large-scale motions, sub-grid scale models and further grid refinement have no significant effect on their LES predictions. Lygren and Andersson [LYG04] and Andersson and Lygren [AND06] performed LES of the axisymmetric and statistically steady turbulent flow in an angular section of an unshrouded rotor-stator cavity for Reynolds numbers up to $Re_\Omega = 1.6 \times 10^6$. Their study showed that the mixed dynamic subgrid scale model of Vreman *et al.* [VRE94] provided better overall results compared to the dynamic subgrid scale model of Lilly [LIL92]. Séverac and Serre [SEV07b] proposed a LES approach based on a spectral vanishing viscosity (SVV) technique, which was shown to provide very satisfactory results with respect to experimental measurements for an enclosed rotor-stator cavity and Reynolds numbers up to $Re_\Omega = 10^6$ [SEV07a]. This alternative LES formulation used a spectral approximation based on a modification of the Navier–Stokes equations. The SVV model only dissipates the short length scales, a feature which is reminiscent of LES models, and keeps the spectral convergence of the error. Viazzo *et al.* [VIA12] performed calculations using fourth-order compact finite-difference schemes with a classical dynamic Smagorinsky model as subgrid-scale stress for $Re_\Omega = 4 \times 10^5$ in the same configuration as the one considered by Séverac *et al.* [SEV07a]. Their results showed that both LES methods are able to accurately describe the unsteady flow structures and to satisfactorily predict mean velocities as well as Reynolds stress tensor components. A slight advantage is given to the spectral SVV approach in terms of accuracy and CPU cost. The strong improvements obtained in the LES results with respect to RANS results confirm that LES is the appropriate level of modelling for enclosed rotating flows in which fully turbulent and transition regimes are involved.

The LES results of Séverac *et al.* [SEV07a] were extended in the non isothermal case by Poncet and Serre [PON09a] using the Boussinesq approximation for Rayleigh numbers up to $10^8$. These authors showed in particular that the turbulent Prandtl number is not constant within the boundary layers and that the classical value of 0.9 used by all the RANS models is valid only at a given location within the boundary layers. Tuliska-Snitzko *et al.* [TUL09] developed an algorithm for LES, with subgrid modelling based on the spectral Chebyshev–Fourier approximation. They used a dynamic Smagorinsky eddy viscosity model in which the Smagorinsky coefficient is averaged along the particle trajectories. It has the advantage that it is applicable also to inhomegeneous flows and that it allows performing computations for higher Reynolds number flows, which are strongly inhomogeneous and anisotropic due to the combined effects of rotation and confinement. It was applied to 3D turbulent and transitional non-isothermal flows within a rotor-stator cavity using the Boussinesq model.

### 1.2.4. Heat transfer modeling

Since the early 1980s, a particular effort has been done to develop innovative RANS models capable of predicting accurately the mean and turbulent flow fields under isothermal conditions. On the contrary, heat transfers have been addressed only with relatively simple approaches. The first reason may be due to the lack of reliable experimental databases, the acquisition of heat transfer information being rather complex and expensive in such systems. Thus, it has slow down the development of specific turbulence models for heat transfer predictions. The second reason is that very simple approaches, such as the one used by Iacovides and Chew [IAC93a], Poncet and Schiestel [PON07] and Manceau et *al.* [MAN09] based on a gradient hypothesis with a given turbulent Prandtl number, have proved sufficient to provide satisfactory results for the Nusselt number distributions. Poncet and Schiestel [PON07] compared indeed very favourably the predictions of the RSM of Elena and Schiestel [ES96] with the experimental data of Sparrow and Goldstein [SPA76] in the case of an impinging jet in a confined rotor-stator cavity considering a large temperature difference (75 K) between the incoming fluid and the walls. Some works using commercial CFD codes provided also reasonable results. Vadvadgi and Yavuzkurt [VAD10] used ANSYS/FLUENT to numerically simulate the flow and the heat transfer in a rotor-stator system of aspect ratio *G*=0.1 and a Reynolds number equal to one million. The simulation was performed using the RSM model with a conjugate heat transfer approach. Among the few advanced heat transfer models which have been developed, we can cite the one of Abe *et al*. [ABE96], who proposed an improved version of the model developed by Nagano *et al.* [NAG91]. It consists of a two-equation heat transfer model, which incorporates essential features of second-order modelling. They introduced the Kolmogorov velocity scale, instead of the friction velocity, to take into account the low Reynolds number effects in the near-wall region and also complex heat transfer fields with flow separation and reattachment. This model has not been yet implemented for rotor–stator flows. We can also mention Craft [CRA91], who has developed advanced transport closures for the turbulent heat fluxes.

## 2. Geometries and results

### 2.1. Free disk

Rotating disk in quiescent air is a basic configuration presented on Figure 1. It has been studied for long years.

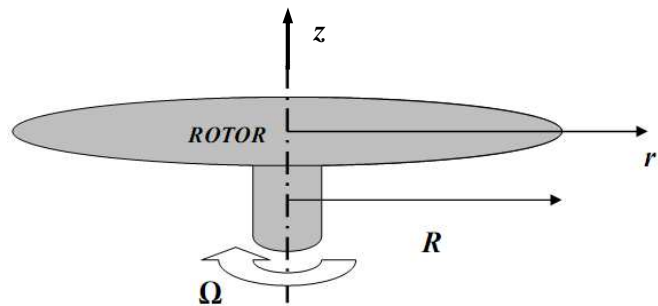

Figure 1. Schematic diagram of a free rotating disk

Very detailed information can be found in [OWE89] and [SHE09a]. Von Kármán [VON21] was the first author to describe the flow and identified the boundary layer developing near the single rotating disk, whose thickness depends on the rotational velocity and the fluid kinematic viscosity. The two main velocity components playing an important role here are the radial and tangential velocities. Outside this boundary layer, only a non-zero axial velocity component can be measured. This flow structure is schematically depicted in Figure 2. The tangential velocity component *v(z)* equals to Ω*r* on the wall and thus increases with the increasing radius being always zero outside of the boundary layer; the radial velocity component *u(z)* exhibits a profile typical wall jets being zero both on the wall ad outside of the boundary layer and demonstrating a maximum (increasing with increasing radius *r*) in the vicinity of the wall; the axial velocity component *w(z)* is uniform and directed towards the surface of the rotating disk, which thus performs as a centrifugal pump, sucking air in the axial direction and pumping it out in the radial direction.

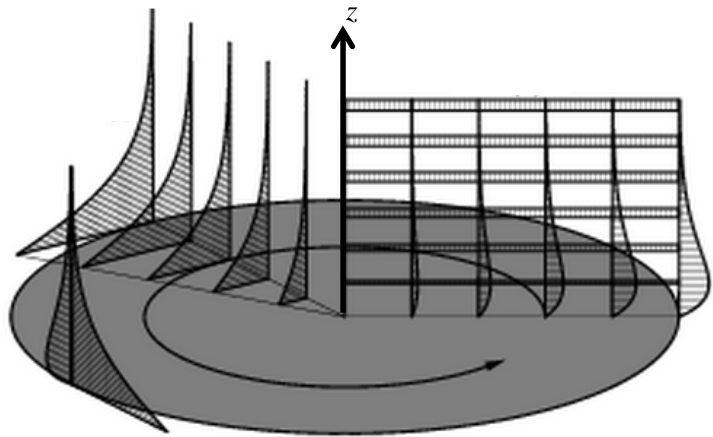

Figure 2. Flow structure over a rotating disk in still air.

The self-similar solutions of the Navier-Stokes and energy equations in laminar flow is well-documented and its results are presented on Figure 3 in comparison with experimental data of different authors, where $u^* = u/\Omega r$; $v^* = v/\Omega r$; $w^* = w/\sqrt{\nu\Omega}$ and $z^* = z\sqrt{\Omega/\nu}$. As can bee seen from Figure 3, agreement of the self-similar solutions with experimental data is excellent.

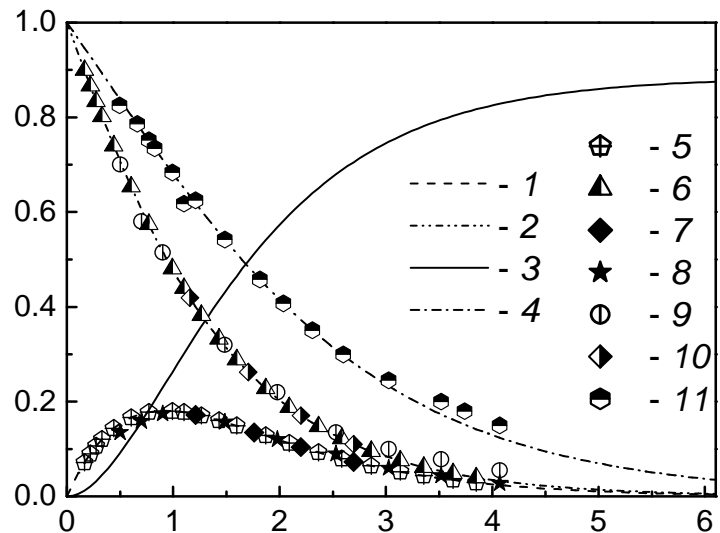

**Fig. 3. Velocity and temperature profiles in laminar flow over a free rotating disk [SHE09a]. Computations [SHE09a]: *1 – F, 2 – G, 3 – –H, 4 –* $\theta = (T - T_\infty)/(T_w - T_\infty)$ (subscript „∞" relates to the flow outside of the boundary layer) for *an isothermal disk ($T_w$=const), Pr=0.71*. Experiments: *5 – F* [ITO94], *6 – F* [ELK97], *7 – F* [LIN96], *8 – G* [ITO94], *9 – G***

**[ELK97]**, *10 – G* **[LIN96]**, *11 – θ for T<sub>w</sub>=const, Pr=0.71* **[ELK97]**.

The nature of the flow over a rotating disk is strongly linked with the local Reynolds number $Re_r=\Omega r^2/\nu$. The flow becomes unstable when the rotational velocity and/or radius over the disk are high enough. For the entire disk, this means there is always a laminar flow region around the rotation axis; in other words, the flow cannot be turbulent over the entire disk. As pointed out in the review made in [SHE09a], the range of the local Reynolds number where the flow undergoes transition from laminar to turbulent regime is not well-established and varies a lot depending on the experimental apparatus and measurement techniques used (see also in [NOR88]). Different authors used different criteria for determining the critical value of the Reynolds number. According to the review [SHE09a], the lowest value of the Reynolds number for the beginning of transition is $1.8 \times 10^5$ [TIE63] while the highest one for the end of transition is $3.6 \times 10^5$ [ELK97]. The emergence of spiral vortices in the near-wall flow is characteristic of the transition to the turbulence and their number can be related to the local Reynolds number by the equation [MAL81]:

$$n_v = 0.0698 Re_R^{0.5} \tag{10}$$

Convective heat transfer on a rotating disk has been studied by many authors, for instance, [GOL35, WAG48, MIL52, HAR59, RIC63, DOR63, KRE68, DEV75, POP75, OEL78, CAR96, OWE89, ELK97, SHE09a]. They have highlighted the fact the heat transfer on the surface of the disk strongly depends on the radial temperature profile. Authors usually correlate the local and mean Nusselt numbers by power laws such as:

$$Nu_r = a\, Re_r^b \tag{11}$$

$$\overline{Nu_R} = c\, Re_R^d \tag{12}$$

The coefficients *a* and *c* depend on the flow regime, Prandtl number and the radial temperature distribution on the disk, while exponents b and d depend, in general, on the flow regime (in other words, the range of the Reynolds numbers).

### 2.1.1. Laminar flow

When the local Reynolds number $Re_r$ is lower than $1.8 \times 10^5$-$3.6 \times 10^5$ [SHE09a], the flow is laminar. In this case, the kinematic boundary layer has a constant thickness estimated from the self-similar solution as $\delta = 5.5\sqrt{\nu/\Omega}$ [OWE89]. As a result, the local convective heat transfer coefficient is independent of the radial position and increases with the rotational speed as the boundary layer becomes thinner. The local Nusselt number given by Eq. (10) varies also with r, with the exponent b being equal to 0.5.

The mean Nusselt number defined by Eq. (11) is constant, with a=c and d=0.5. In other words, for laminar flow, the mean Nusselt number is by definition equal to the local Nusselt number calculated at the radius where the laminar flow ends (in case where a turbulent flow also develops) or at the external radius (in case of a laminar flow over the entire disk), denoted $R_t$:

$$\overline{Nu_{R_t}} = a\, Re_{R_t}^{0.5} \tag{13}$$

In the case of an isothermal surface, the experimentally/theoretically determined coefficient *a* varies for air depending on the author: $a = 0.28$ [MIL52], $a = 0.33$ [HAR59, POP75], a≈0.32 [ELK97], $a = 0.335$ [WAG48], $a = 0.341$ [OEL78], $a = 0.345$ [KRE68], *a*=0.36 [COB56], $a = 0.38$ [GOL35], $0.322 < a < 0.343$ [DOR63]. Differences can be attributed to inaccuracies of measurement and also to possible differences in surface roughness [SHE09a]. The exact self-similar solution of the boundary layer equations yields $a = 0.3286$ for $Pr = 0.72$ [SHE09a].

In the case of a radial distribution of temperature which can be interpolated by a law expressed as $T(r) = T_0 + \beta r^n$, Dorfman [DOR63] and Owen & Rogers [OWE89] obtained a semi-empirical solution, which for air at Pr=0.72, writes:

$$a = 0.261\sqrt{(n+2)} \tag{14}$$

As shown by [SHE09a] (see also in Table 1), this equation agrees quite well with the self-similar solution for n≥2, however deviates from it for n<2, with these deviations increasing for the decreasing n, and being already noticeable for an isothermal surface (n=0) and becoming very significant for negative values of n.

An alternative approximate solution was obtained by [SHE01, SHE09a] for a wide range of the Prandtl number; for Pr=0.72, this solutions looks as:

$$a = \frac{0.4435}{0.3486 + 2.002/(2+n)} \qquad (15)$$

The values of the coefficient *a* for Eq. (10) computed by Eq. (14) (column 3, Table 1) agree very well with the exact self-similar solution (column 2, Table 1) over the entire range of the values of n=[-1.5; 4].

| n | Exact solution [SHE09a] | Equation (14) [SHE09a] | Equation (13) [DOR63] | Oehlbeck [OEL78] | Hartnett [HAR59] |
|---|---|---|---|---|---|
| -1.5 | 0.1045 | 0.1019 | 0.187 | - | - |
| -1.0 | 0.1911 | 0.1887 | 0.265 | - | - |
| -0.5 | 0.2647 | 0.2635 | 0.326 | - | - |
| 0 | 0.3286 | 0.3286 | 0.374 | 0.341 | 0.33 |
| 1 | 0.4352 | 0.4365 | 0.459 | 0.436 | 0.437 |
| 2 | 0.5223 | 0.5223 | 0.53 | 0.519 | 0.512 |
| 4 | 0.6599 | 0.6500 | 0.649 | 0.573 | 0.661 |

**Table 1. Averaged Nusselt numbers over a rotating disk for various temperature differences.**

[OEL78] and [HAR59] obtained values of the coefficient *a* also presented in Table 1. In general, all values for n = 0 are obviously concordant with the experimental data for an isothermal surface mentioned above.

Fundamental peculiarities of the effects of the disk permeability in laminar flow being of perspective interest for the rotor/stator flows with a perforated stator were studied in [TUR11a, TUR11b].

### 2.1.2. Transition from laminar to turbulent flow

The transition of the flow from laminar to turbulent flow regime is very dependent on the experimental conditions as demonstrated in [NOR88, SHE09a]. Several empirical equations for the local heat/mass transfer for transitional flow are documented in [SHE09a]. Table 2 represents empirical equations for heat transfer in air.

| Author | Correlation | Lowest Re | Highest Re |
|---|---|---|---|
| [POP75] | $Nu_r = 10.0 \times 10^{-20} Re_r^4$ | $1.95 \times 10^5$ | $2.5 \times 10^5$ |
| [ELK97] | $Nu_r = 2.65 \times 10^{-20} Re_r^4$ | $2.9 \times 10^5$ | $3.6 \times 10^5$ |
| [CAR96] | $Nu_r = 8.01 \times 10^{-14} Re_r^{2.8}$ | $2.6 \times 10^5$ | $3.5 \times 10^5$ |

**Table 2. Local Nusselt numbers over a rotating disk for transitional flows.**

### 2.1.3. Turbulent flow

For a local Reynolds number higher than $2.5 \times 10^5 - 3.6 \times 10^5$ [SHE09a], the flow over the disk becomes turbulent. Looking at the entire rotating disk, turbulent flow region is located in the outer peripheral area (Fig. 4). The boundary layer thickness in turbulent flow can be estimated from the equation derived using the integral method by von Kármán $\delta = 0.526\, r^{1.8} \left(\nu/\Omega\right)^{0.2}$ [VON21]. More detailed information on the relations for different parameters of the boundary layer over a rotating disk obtained using an integral method is documented in the monograph [SHE09a].

In turbulent flow, the local heat transfer coefficient, unlike that in the laminar flow, increases with both the radius and rotational speed. The exponents *b* and *d* in Eqs. (10) and (11) equal to 0.8 or are close to this value. Like in the laminar flow case, the radial temperature profile over the disk surface affects noticeably the coefficients *a* and *c* in Eqs. (10) and (11).

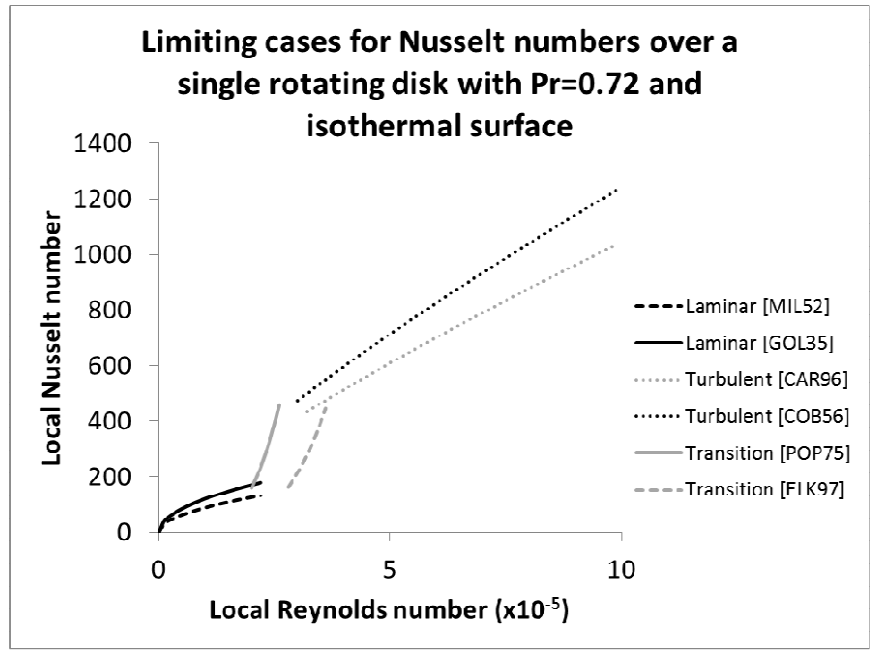

Figure 4. Local Nusselt numbers over a rotating disk in quiescent air for isothermal surface.

For $Pr = 0.72$ and a radial temperature distribution approximated by $T(r) = T_0 + \beta r^n$ and, Dorfman [DOR63] and Owen & Rogers [OWE89] obtained also a semi-empirical solution for the coefficients in Eqs. (10) and (11) for turbulent flow:

$$a = 0.0162(n + 2.6)^{0.2} \qquad (16)$$

$$c = 0.0162 \frac{n+2}{(n+2.6)^{0.8}} \qquad (17)$$

For an isothermal disk the values of these coefficients in accordance with Eqs. (15) and (16) are *a*=0.0196 and *c*=0.0151.

Having experimentally investigated heat transfer of an isothermal disk with air (*n*=0, Pr=0.72), the authors of [COB56, NIK63] determined the values of the coefficient *a* in Eq. (10) equal to 0.0197 and 0.0194, respectively. Experimental values of the coefficient *c* found by [BUZ66, KUZ62, COB56, DEN70] are equal to 0.15. De Vere [DEV75] also came to Eq. (11) with *c*=0.015. These experimental data agree with the estimates in accordance with Eqs. (15) and (16). However, experimental data [MCC70] with a=0.0179, and especially more recent [POP75] with a=0.0188 and [ELK97] with a=0.0189 testify in favour of lower values of *a* than those predicted by Eq. (15). Especially salient becomes this tendency for the boundary condition on the disk surface with $q_w$=*const*, where Eq. (15) predicts the value a=0.0186, while the recent and very accurate experimental data [ELK97, CAR96] again indicated a lower value of a=0.0163. Numerical simulations [TAD83] for the boundary condition $q_w$=*const* are generalized by Eqs. (10) with *b*=0.83 and *a*=0.0111; this curve lies below the estimate of Eq. (15) and close to the experiments of [ELK97, CAR96].

An improvement of the agreement of the theory and experiments provides an alternative approximate solution derived by [SHE00, SHE09a] and valid for a wide range of the Prandtl numbers; for Pr=0.72, this solutions looks as:

$$a = \frac{1}{34.99 + \dfrac{48.33}{2.6+n}} \qquad (18)$$

For the isothermal disk (n=0), this solution yields a=0.0187, whilst for $q_w$=*const* it results in a=0.0169; unlike the Dorfman's solution, Eq. (15), both these values are in excellent agreement with the aforementioned experimental/theoretical data of [MCC70, POP75, ELK97, CAR96, TAD83].

As mentioned above, both exact and approximate analytical solutions of the thermal boundary equations in form of the Eqs. (10), (11) were possible for the boundary condition $T(r) = T_0 + \beta r^n$. A new type of the wall boundary condition providing a possibility of an analytical solution of the thermal boundary layer equation for a free rotating disk was revealed and discussed in details in [SHE05, SHE09a]. This solution represents a two-parameter family of curves that allows modeling wall temperature profiles with maxima and minima and values of *d* noticeably different from 0.8. As mentioned in [SHE05, SHE09a], this solution provides much better accuracy of the agreement with the experiments [NOR88] than the Eqs. (10), (15) and (16), which significantly strengthens the analytical toolkit for modeling heat transfer of a rotating disk.

### 2.2. Disk in cross flow

Rotating disks in quiescent air were elucidated very detailed in the literature. However, the configuration with a cross flow parallel to the plane of rotation was paid far less attention. This configuration is presented on Figure 5. The configurations with an inclined rotating disk (with an angle of attack to the incident flow lying with the range of 0 to 90 deg) were first studied only very recently [WIE10, TRI11].

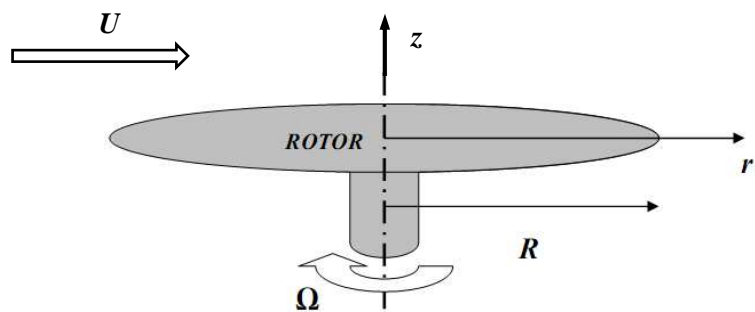

Figure 5. Schematic diagram of a rotating disk in cross flow

### 2.2.1. Stationary disk

In the case of a stationary disk in air crossflow, aus der Wiesche [WIE07] numerically studied local convective heat transfer for air crossflow Reynolds numbers $Re_U$ (based on incident flow velocity U and external radius R) varying from $10^3$ to $10^6$. Good agreement is obtained between its results obtained by LES and those for the flat plate, characterized by the symmetric Nusselt number (Fig. 6a). Locally, a high convective heat transfer occurs at the beginning of the disk, in the developing boundary layer, and decreases along the disk diameter. The flow is laminar up to a critical value of $Re_U = 5 \times 10^4$, for which the transition into turbulence begins, whereas the turbulent regime is fully reached for $Re_U > 10^5$. The LES data are correlated by Eq. (18) for the laminar case, and by Eq. (19) for the turbulent regime.

$$\overline{Nu_R} = 0.417 Re_U^{0.5} \tag{19}$$

$$\overline{Nu_R} = 0.0127 Re_U^{0.8} \tag{20}$$

Dennis et al. [DEN70], based on their original experimental data for turbulent flow, developed an empirical correlation which yields significantly higher mean Nusselt numbers than those suggested by Eq. (19):

$$\overline{Nu_R} = 0.027 Re_U^{0.8} \tag{21}$$

The theoretical solution presented also by Dennis et al. [DEN70] suggested the constant in Eq. (20) to be 0.027 rather than 0.036.

Measurements by He et al. [HE05] were described by the empirical equation:

$$\overline{Nu_R} = 0.0058 8 Re_U^{0.925} \tag{22}$$

which is concordant with the aforementioned theoretical solution of Dennis et al. [DEN70] and still suggests significantly higher Nusselt number than Eq. (19).

As pointed out in the review [SHE09a] analysing the given above equation, the reason for the Eq. (19) to suggest much lower values of the Nusselt numbers than those measured by [DEN70, HE05] is that the disk simulated by aus der Wiesche [WIE07] had a zero thickness, which thus prevented formation of the flow separation and reattachment region close to the leading edge of the disk and subsequent significant heat transfer enhancement.

In the recent experimental works [WIE10, TRI11], it was also shown that the LES results by [WIE07] underestimate the real heat transfer for such configuration. Authors also point out the influence of the thickness of the disk when positioned in a cross flow. Indeed, the disk thickness affects the size of the separation zone that can strongly influence the local heat transfer in the part of the disk close to its leading edge.

### 2.2.2. Rotating disk

In the case of a rotating disk in air cross-flow, few studies are known that deal with the combined effect of rotation and air cross-flow on convective heat transfer. The most detailed theoretical work was performed by aus der Wiesche [WIE07] who numerically studied the flow field and the corresponding local Nusselt number distributions on the disk for $10^3 < Re_\Omega < 10^6$ and $10^3 < Re_U < 10^6$. In this study, the flow is parallel to the surface of the rotating disk that has zero thickness. By comparison between the results obtained with and without disk rotation, aus der Wiesche [WIE07] analysed the rotation effect on convective heat transfer. For high values of the cross-flow Reynolds, convective heat transfer is governed only by the cross-flow since no additional heat transfer augmentation due to rotation was observed. For low rotational and cross-flow Reynolds numbers, disk rotation affects slightly the flow field and local temperature distribution, whilst the mean heat transfer over the disk surface remains constant. Indeed, rotation increases heat convection on the co-moving side and diminishes it on the counter-moving side in such a way that the mean heat transfer is almost unaffected. While increasing $Re_\Omega$, the increase of the flow turbulence leads to higher variations of the local Nusselt number, even if the typical flat plate distribution can still be identified. For the sufficiently high values of $Re_\Omega$, the local Nusselt number distribution becomes uniform and mainly governed by rotational effects (Fig. 6b).

It was shown that the air cross flow has the dominant influence for the mean heat transfer while $Re_\Omega < 1.4 Re_U$. In fact, the mean Nusselt number remains constant at the value obtained without rotation. In the second zone, corresponding to $Re_\Omega > 1.4 Re_U$, there is a significant increase in the mean Nusselt number due to some instabilities, created between rotational and plane flows, which tend to increase the convective heat transfer. Based on the LES study, the author proposes different correlations in order to predict convective transfer on the disk.

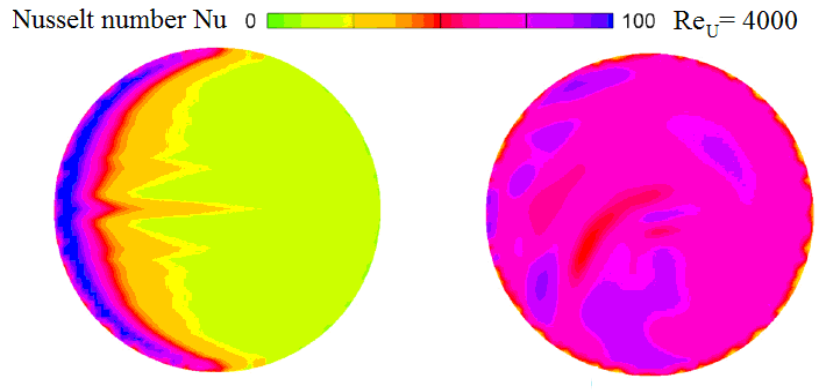

Figure 6. Local Nusselt numbers over a rotating disk in cross flow
[WIE07]

By comparison with the mean Nusselt number without rotation, two correlations are proposed:

for $0 < Re_\Omega < 1.4\, Re_U$:

$$\overline{Nu} = \overline{Nu_{\Omega=0}} \tag{23}$$

and for $1.4\, Re_U < Re_\Omega < 5\, Re_U$:

$$\overline{Nu} = \overline{Nu_{\Omega=0}}\left[1 + 0.32\left(\frac{Re_\Omega}{Re_U} - 1.4\right)^{0.5}\right] \tag{24}$$

They can also be calculated with:

$$\overline{Nu} = \sqrt{(0.0127\, Re_U^{0.8})^2 + (0.015\, Re_\Omega^{0.8})^2} \tag{25}$$

in the case where $Re_\Omega > 2 \times 10^5$ and $Re_U > 5 \times 10^4$, or by:

$$\overline{Nu} = \sqrt{(0.0127\, Re_U^{0.8})^2 + (0.33\, Re_\Omega^{0.5})^2} \tag{26}$$

in the case where $Re_\Omega < 2 \times 10^5$ and $Re_U > 5 \times 10^4$.

Being useful to understand the nature of mutual interaction of the rotation and cross-flow effects, Eqs. (22)-(25) obtained for a zero thickness disk can not be applied for computation of heat transfer over a disk of a finite thickness.

In the case of a fixed disk mounted on a cylinder and submitted to an air cross-flow, the presence of the cylinder generates flow perturbations, at the disk/cylinder junction, studied by different authors [GOL84, SUN96, SCH87, SCH92, SPA86, FIS90]. From experimental observations, they detected a boundary layer development from the leading edge of the disk associated to a reduction of velocity due to the adverse pressure gradient in the stagnation zone upstream of the cylinder. This causes the flow to separate and to form a horseshoe vortex system (Fig. 7), consisting of counter-rotating vortices swept around the cylinder base. Zones of lower convective heat transfer correspond to the wake and to the flow separation located at θ = ± 90° from the front stagnation point. On the other hand, zones of higher heat transfer are located at +110°<θ<+150° and -110°<θ<-150°, where the horseshoe vortex system appears. Experimental studies [ROU05, FU05, SAH08] dealing with the visualization of the horseshoe vortex system by Particle Image Velocimetry highlighted a significant influence of $Re_U$ on the size of the vortex horseshoe. Moreover, Fu and Rockwell [FU05] showed that the instability of the horseshoe vortex generates flow perturbations in the near wake, fluctuation level increasing in the horseshoe vortex leads to the development of coherent vortices earlier in the separating shear layer, and rotational perturbation of a vertical cylinder can destabilize the wake flow structure.

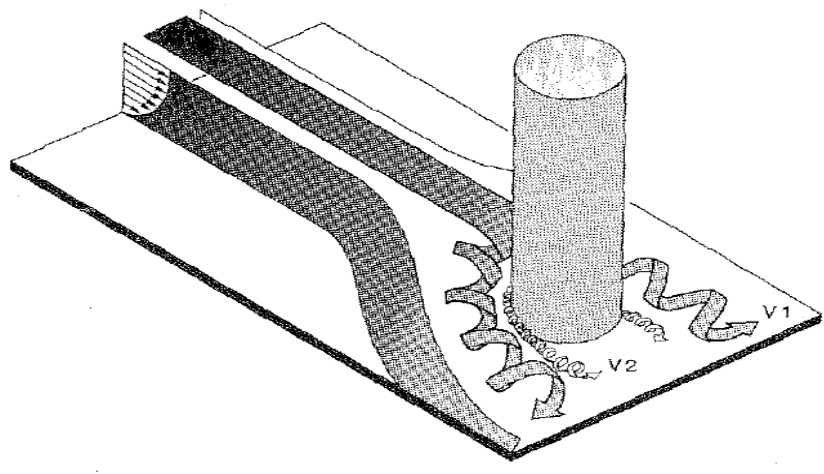

Figure 7. 3D boundary layer separation and horseshoe vortex system in the region of interaction between mainstream boundary layer and cylinder [GOL84]

To the best of our knowledge, there is only one previous study about the mean convective exchanges around a rotating disk mounted on a cylinder in airflow [WAT97]. In the case of a disk of 21-mm height and 1-mm wide mounted on a 58-mm diameter cylinder, Watel [WAT97] developed a general correlation of the mean Nusselt number depending on $Re_U$ and $Re_\Omega$:

$$\overline{Nu} = (0.03\, Gr + 0.053\, Re_U^2 + 9.1 \times 10^3\, Re_\Omega^2)^{0.275} \tag{27}$$

From this equation, different domains of influence (rotation, air cross-flow, combined rotation/air cross-flow) on convective heat transfer were determined.

### 2.3. Jet on stationary and rotating disks

#### 2.3.1. Fluid flow

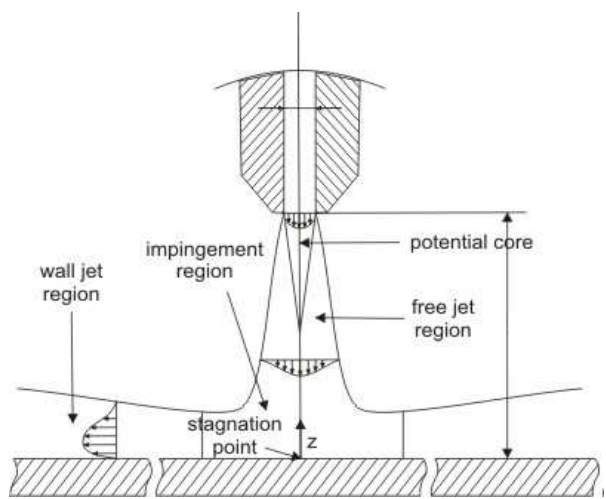

Figure 8. Flow at a jet impingement

**Stationary impinged surface.** Figure 8 shows the flow which develops in a jet on a stationary plate where 3 main areas can be identified.

In the first area just after the pipe exit, a potential core and a free jet region can develops depending on the ratio e/D. The free jet region is a mixing region between the fluid in the jet and the fluid outside the jet so that vortices develop as shown by [ANG03] for e/D=4.5. The vortices are believed to not develop when e/D<1. For large enough aspect ratios e/D, the centreline velocity and turbulence intensity keeps quite constant and equal to the exit velocity up to 2.5 diameters from the nozzle [DON04].

The second region is called the impingement (stagnation) region and it is where the flow modifies from an axial flow to a radial flow. [DON04] experimentally investigated by means of PIV and LDA the whole impingement region for 0.5<e/D<8 and $10000 < Re_j = \frac{WD}{v} < 30000$. The impinging point in the center of the impingement spot is called the stagnation point. It was shown that the stagnation point is fixed and aligned with the pipe only for e/D<2.5. If not, this point radially oscillates and is fixed only on time-averaged visualizations. The boundary layer thickness developing on the impinged plate decreases with increasing radial distances from the impinging point, so velocity and turbulences increase.

The third area is the so called the wall jet region where the flow is fully radial. Boundary layer thickness increases and velocities decrease with an increase in radius. Very interesting studies can be found in [GUE05] who focused on the wall jet region in the case where e/D=2 and $Re_j = 35000$ by the use of hot-wire anemometry. Authors give lot of details about velocity and temperature profiles near the impinged wall. [KNO98] experimentally obtained velocities and turbulence field near the impingement of a jet on a stationary surface for different aspect ratios between 1 and 10. It was shown that velocity profiles reach self-similarity within 2.5D. For turbulence, the self-similarity is obtained for dimensionless radii higher than 4.5D. Peak levels of turbulent fluctuations occur at about 2D.

**Rotating impinged surface.** The superposition of the disk rotation and the jet impingement was of interest in [ITO98, MIN04] as far as the flow was concerned. In the case where e/D=5 and $Re_j = 14000$, [MIN04] showed that the flow is unaffected by the rotation if r/D is lower than a **critical** dimensionless radius, which increases as the rotational speed decreases. This phenomenon can be seen in Figure 9. The authors used a parameter $\alpha = \frac{\Omega r^2}{WD}$, which described the ratio of the centrifugal forces and strength of the impingement. [MIN04] identified three areas over the rotating disk: at α<4.7, the flow is unaffected by the rotation, for 4.7<α<9.2 the flow is affected by rotation and by the impinging jet, and finally for α>9.2, the flow is only affected by the rotation.

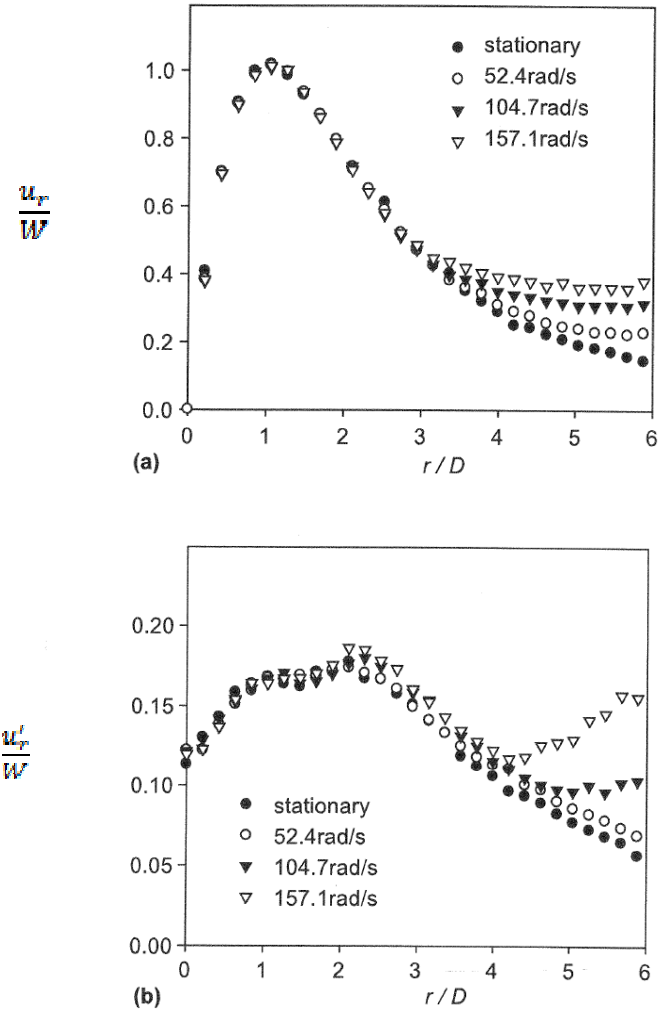

Figure 9. Radial distribution (at z/D=0.032) of the near-wall velocity; (a) mean velocity components; (b) turbulent intensity; $Re_j$=14500 and e/D=5 [MIN04]

### 2.3.2. Heat transfer

**Stationary impinged surface.** Viskanta [VIS93], in his review paper, summarized the studies concerning heat transfer due to single jet on stationary surfaces performed before 1993 and did not focus on rotating systems. Several authors showed that convective heat transfer near the impingement region is strongly influenced by the aspect ratio e/D and jet Reynolds number $Re_j$. In the region r/D<5, one or two peaks in the local Nusselt number distribution were observed.

When e/D>4, the impingement between the jet and the plate is situated at the end of the potential core of the jet, leading to a maximal heat transfer at the stagnation point which monotonically decreases with the radius. For lower aspect ratios, two peaks can eventually be observed in the radial profile of local Nusselt numbers, which are not aligned with the center of the jet. The first peak is located at r/D=0.5 while the position of the secondary peak varies (as shown on Figure 10) according to the relation [LYT94]

$$\left(\frac{r}{D}\right)_{secondary\,peak} = 0.188 Re_j^{0.241} \left(\frac{e}{D}\right)^{0.224} \quad (28)$$

over the range: $11000 < Re_j < 27600$ and $0.1 < \frac{e}{D} < 1$.

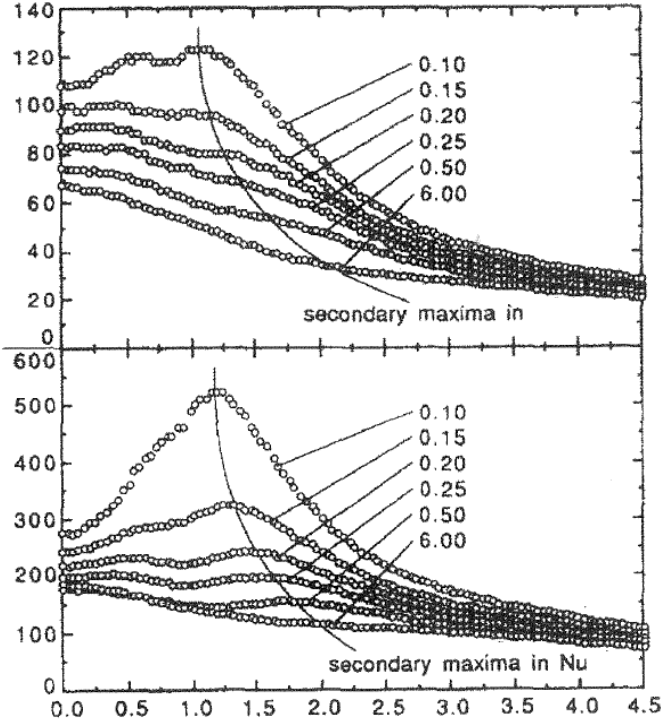

Figure 10. Radial variation in Nusselt number with nozzle-plate spacing for two jet Reynolds numbers on a stationary disk [LYT94]

The work from Lytle and Webb [LYT94] also clearly showed that this secondary peak is due to the transition from laminar to turbulent boundary layers which occurs earlier when the aspect ratio decreases. [DON04] explains this secondary peak by the fact the boundary layer thickness reaches its minimum and that it is mixed with the free jet shear layer which increase turbulence levels at this point.

The local Nusselt number at the stagnation point was also evaluated and correlated by:

$$Nu_D = 0.726\, Re_j^{0.53} \left(e/D\right)^{-0.248} \qquad (29)$$

It is quite difficult to find out correlations of local Nusselt numbers in the impingement region. In our knowledge, only [SAN97] proposed such correlations which are quite complex. Much information on the surface distributions of local Nusselt numbers can be found in the previously cited papers and in [YAN97, LEE02, GAO06, LIU08, KAT08].

**Rotating impinged surface.** Other authors have studied mass and heat transfer variation over a single rotating disk [POP86, OWE89, CHE98]. Popiel and Boguslawski [POP86] classified their experimental results into three categories. The first one is the area of the disk where the influence of the jet on the heat transfer is the most significant. The second one is the area where rotation has the domineering effect. The third one is a mixture of the first two. Chen et al. [CHE98] and Sara et al. [SAR08] confirmed those observations, concluding that the location of these three zones depends on the ratio of the jet and rotation mass flow rates. They also concluded that the heat transfer on a rotating disk is affected by the jet only for the local Reynolds number $Re_r < 200000$. [CHE98] presented observations on the variation of the distance $R_j$ between the center of the jet and the center of the rotating disk. Results are depicted in Figure 11. [CHE98] concluded that *the best jet position is the center of the rotating disk, which is the position that maximizes heat transfer*.

[AST08] studied jet impingement on a rotating disk for high aspect ratios (3<e/D<42.5) and rotational Reynolds numbers under the transitional one. For these conditions of laminar flow, authors correlate the stagnation Nusselt number with the relations:

$$Nu_D = Re_r^{0.5}(0.33 + 1.57\xi) \qquad (30)$$

when $\xi = Re_j^2 \left(\dfrac{v}{\Omega e^2}\right)^{1.5}$ is under 1, or

$$Nu_D = 1.81\, Re_r^{0.5}\, \xi^{0.597} \qquad (31)$$

when the jet effect is assumed to be dominating ($\xi > 1$).

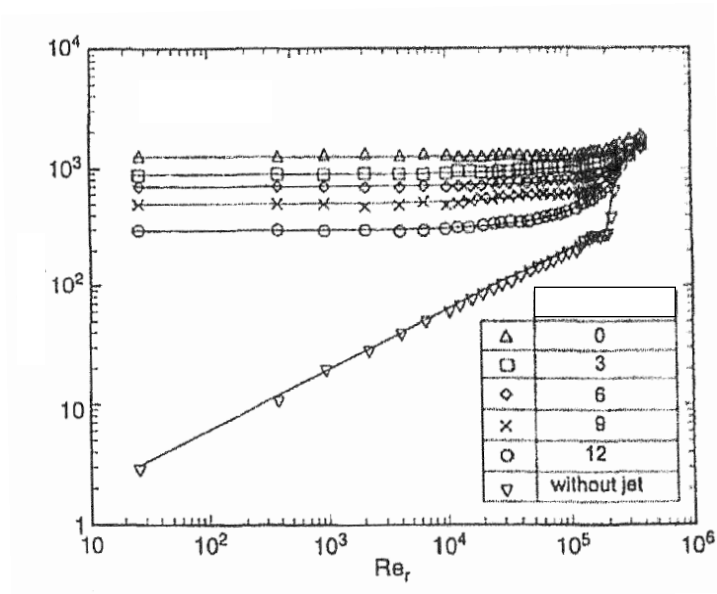

Figure 11: Local heat/mass transfer of rotating disk with impinging jet at e/D=5 [CHE98]

Under this last condition, authors also provided a correlation for the end of the jet dominated regime that pinpoints the radius over the disk surface where results start varying with the variation of the rotational speed:

$$Re_r = 11200 Re_j^{0.24} \tag{32}$$

Unfortunately, the aspect ratio (e /D) effect is not included in this correlation.

Though empirical correlations for the local Nusselt numbers were not developed in [SANI98, SAR08], surface distributions were depicted in these works.

### 2.3.3. Exact solution for the stagnation point

As said above, the best jet position is the center of the rotating disk, because this positioning maximizes the surface heat transfer. Shevchuk et al. [SHE03, SHE09a] following commonly accepted theoretical model assumed the boundary layer in the stagnation region over a rotating disk subjected to uniform outer flow orthogonal to the disk to be fully laminar, with the radial velocity at the outer boundary of the boundary layer increasing linearly as $u_{r,\infty} = ar$. The constant $a$ according to the known exact solution equals to $a = \dfrac{4W}{\pi R}$, where R is the outer radius of the disk. Exact self-similar solution of this problem is tabulated in [SHE09a] for different Prandtl numbers and values of the exponent *n* in the boundary condition for the wall temperature $T(r) = T_0 + \beta r^n$. Introducing the parameter κ=$a/\Omega$ that reflects the relative strength of the impingement flow in comparison with the rotation effects, Shevchuk et al. [SHE03] derived a solution for the Nusselt number in the stagnation point:

$$Nu_{2R} = K_1(1+\kappa^{-1})^{0.5} Re_j^{0.5} A^{0.5} \tag{33}$$

$$K_1 = 0.6159 \chi \frac{(1+\kappa)^{-0.5} Pr \tau_\theta}{\tau_{\theta 0}} \tag{34}$$

$$\frac{\tau_\theta}{\tau_{\theta 0}} = \left[\frac{\alpha + 1.301\kappa}{\alpha_0}\right]^{0.5} \tag{35}$$

$$\alpha = -0.4275\kappa + (2.717\kappa^2 + 0.6863)^{1/2} \tag{36}$$

Here the Reynolds analogy parameter is to be found from the equation

$$\frac{1}{b_2} - 0.6518 \chi Pr^{n_p} \frac{b_1}{b_2} + \frac{\kappa}{\alpha}(e_1 \chi^{-1} + e_2 \chi + e_3) = 0.3482 \cdot \chi \frac{4}{2+n}\left[1+1.301\frac{\kappa}{\alpha}\right] \tag{37}$$

where $e_1$=0.2007, $e_2$=-0.7495, $e_3$=1.3556, $b_1$=0.6827, $b_2$=1.4373 and $n_p$=0.7436 for Pr=0.72; $Re_j = WD/\nu$; $Nu_{2R}$ is based on the disk's diameter 2R as the characteristic length and A=aD/W. $\chi$ is the Reynolds analogy parameter.

As shown in [SHE03, SHE09a], at κ=$a/\omega$>1.5 heat transfer is dominated only by peculiarities of the impinging jet. Detailed comparisons with the cases of the uniform orthogonal flow onto a rotating disk and co-axial impingement of round jets of the diameter much smaller than the disk's diameter performed in [SHE03, SHE09a] showed good efficiency of the above solution in simulations of such geometrical configurations.

### 2.4. Rotor-Stator without jet

In this section, the review is restricted to the smooth disk case in small aspect ratio cavities. High aspect ratio may indeed favour the development of the vortex breakdown phenomenon, which is not discussed here.

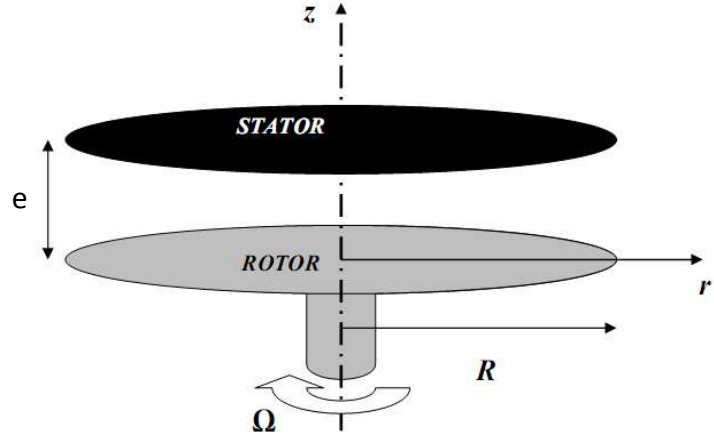

Figure 12: Rotor-stator configuration without impinging jet.

We consider the flow between a rotating disk (the rotor) facing a stationary one (the stator) as it can be seen in Fig. 12. Both disks have the same outer radius *R* and are separated by an axial gap denoted *e* in the following. The cavity may be enclosed or not by a fixed external cylinder (the shroud). The influence of a superimposed throughflow in an opened system will be discussed in the next section. An inner rotating cylinder (the hub) may be attached (annular cavity) or not (cylindrical cavity) to the rotor. Thus, both the hub and the rotor rotate at the same angular rotation rate $\Omega$. The main effect of finite radial extension of the disk and most of all of the inner and outer cylinders is that the boundary conditions are not compatible with the self-similarity solutions of Ekman [EKM02], Von Kármán [VON21] or Bödewadt [BOD40] though there may be qualitative resemblance far from the end walls. The reader can refer to the detailed monograph of Owen and Rogers [OWE89] for an overview of the experimental, theoretical and numerical works until 1989 including heat transfer effects and to the recent review paper of Launder *et al.* [LAU10] for a detailed review of isothermal rotor-stator flows in enclosed cavities.

### 2.4.1 Base flows

The first major experimental and theoretical study of enclosed rotor-stator flows was carried out by Daily & Nece [DAI60], who identified the four flow regimes shown in Fig. 13. The base flow is indeed governed by two global parameters: the aspect ratio of the cavity *G=e/R* and the rotational Reynolds number $Re_\Omega = \Omega R^2 / \nu$. The flow is then either laminar (regimes I & II) or turbulent (regimes III & IV) with merged (I & III) or unmerged (II & IV) boundary layers. Some results may be also discussed in terms of only one global parameter: the Reynolds number based on the interdisk spacing defined as $Re_e = G^2 \cdot Re_\Omega$. In all flow regimes, the base flow is characterized by a strong circulation in the tangential direction and a secondary flow in the meridian plane. The main difference between merged and unmerged boundary layer flow regimes consists only in this secondary flow as we will see below.

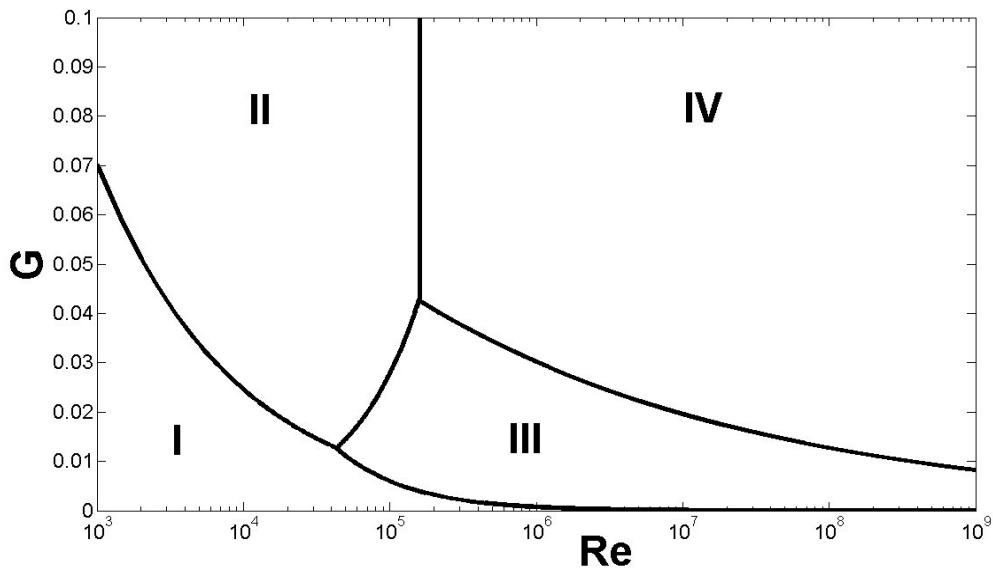

Figure 13: The four flow regimes in an enclosed rotor-stator cavity after Daily & Nece ([DAI60]. Merged boundary layers: I (laminar) and III (turbulent). Unmerged boundary layers: II (laminar) and IV (turbulent).

In regimes II and IV, the secondary flow structure in the meridian plane is known as the Batchelor flow structure since the similarity analysis of Batchelor [BAT51], who solved the system of differential equations relating to the stationary axisymmetric flow between two infinite disks. At a given radius *r*, the velocity profiles are divided into three distinct regions: two boundary layers develop on each disk separated by a non viscous core region (Figure 14a to c). It can be seen as the connection of the Bödewadt and Ekman flow problems over single infinite disks. Thus, by analogy with these prototype flows, the boundary layers along the rotor and the stator are known as the Ekman and Bödewadt layers respectively. The geostrophic core is characterized by quasi zero axial and radial velocities and a constant tangential one. In its normalized form, the tangential velocity in the core of the flow is known in the literature as either the core-swirl ratio $\beta$ or the entrainment coefficient $K$ (=$V_\vartheta/(\Omega r)$, defined as the ratio between the tangential velocity of the fluid in the core and that of the disk as the same radius). As a consequence of the Taylor-Proudman theorem, there is also no axial gradient in that region. Thus, one can readily link the core-swirl ratio $\beta$ to the radial pressure gradient at the rotor, which is crucial for designing thrust bearings. In both regimes II and IV, the fluid is pumped centrifugally outwards along the rotor and is deflected in the axial direction after impingement on the external cylinder (Fig. 15). After a second impingement on the stator, it flows radially inwards along the stator, by conservation of mass, before turning along the hub and being centrifuged again by the rotating disk. This secondary flow picture is modified in the absence of the hub [PER11]. The main difference between these two flow regimes mainly consist in the values of the coefficient $\beta$ as we will see in the following.

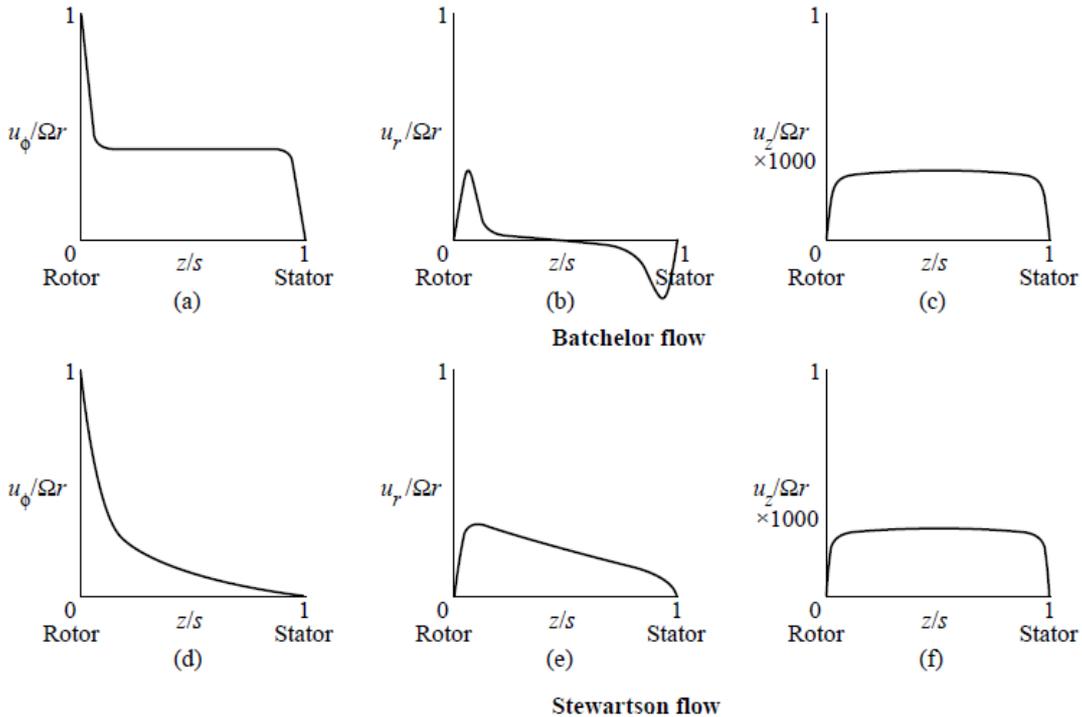

Figure 14: Sketches of the characteristic velocity profiles in a rotor-stator system: (a) to (c) Batchelor flow; (d) to (f) Stewartson flow, after Childs [CHI11].

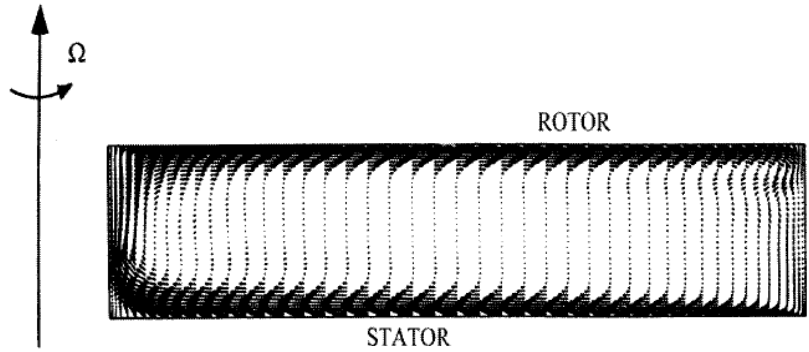

Figure 15: Base flow of Batchelor type with both Ekman and Bödewadt layers separated by a solid body rotating core at $Re_\Omega$ = 15.000 and $G$=0.127. Velocity field in a meridianal plane obtained by Serre *et al.* [SER02] using direct numerical simulation.

In the regimes II and IV, the division of the flow into three distinct zones was the subject of an intense controversy: Stewartson [STE53] found indeed that the tangential velocity of the fluid can be zero everywhere apart from the rotor boundary layer. The radial velocity is also modified with a purely centrifugal outflow in the wheelspace (Fig. 14d to f). Kreiss & Parter [KREI83] finally proved the existence of a multiple class of solutions discovered numerically by Mellor *et al.* [MEL68]. Zandbergen and Dijkstra [ZAND87] also showed that the similarity equations do not generally have unique solutions and that the two solutions advocated by Batchelor and Stewartson can thus both be found from the similarity solutions. In the case of a laminar flow between two finite disks, Brady and Durlofsky [BRA87] found that flows in an enclosed cavity are of Batchelor type, while open-end flows are of Stewartson type. Gori [GOR85] studied the transition between these two solutions for laminar flows in an enclosed cavity. The transition appears either by decreasing the rotational Reynolds number or increasing the aspect ratio of the cavity. A superimposed outward throughflow can favour the transition to the Stewartson structure at large flowrates [PON05b].

Up to now, regimes I and III with merged boundary layers have received less attention due to the very small gap between the disks, which make the measurements difficult to perform with accuracy. When $Re_e$ is sufficiently small, the flow is referred as being of torsional Couette type. The viscous region fills the whole gap and inertial effects may be neglected. Randriamampianina *et al.* [RAND97] performed direct numerical simulations and turbulence modelings for $G$=1/11 and various Reynolds numbers covering the regimes I, II and IV of Daily and Nece [DAI60]. In regime I, they found that the flow is dominated by diffusive motion, and the recirculation cell is centered towards the shroud. The axial variation of the radial velocity component has a quasi-symmetrical behaviour between the stator and the rotor (Fig. 16a), typical of Couette flow. However, a slight non-symmetry, resulting from the different nature of the flow over rotating and stationary discs, is observed. The tangential velocity component at mid-height between the disks is equal to half the local disk speed. This value is constant whatever the radial location, apart very close to the inner and outer cylinders (Fig. 16b). This picture is not significantly modified in regime III [HAD08].

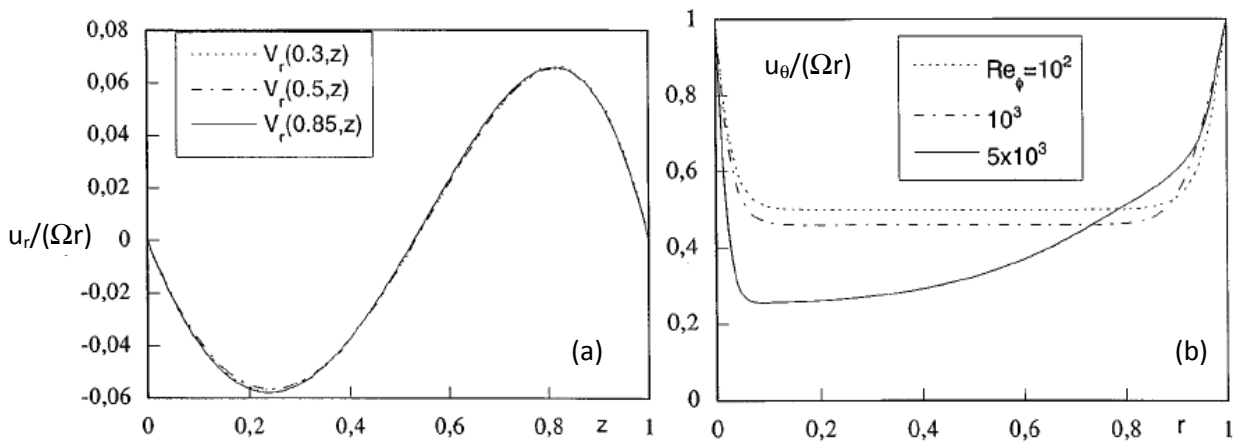

Figure 16: (a) Axial variation of the normalized radial velocity component $u_r/(\Omega r)$ at three radial locations $r/R$ 0.3, 0.5 and 0.85 for $Re_\Omega=10^3$; (b) Radial variation of the normalized azimuthal velocity component $u_\theta/(\Omega r)$ at mid-height of the cavity for three rotational Reynolds numbers. Regime I, after Radriamampianina et al. [RAND97].

For intermediate values of aspect ratio, when one moves from the axis of rotation to the periphery of the cavity (increasing values of the local radius $r$), the boundary layers thicken and turbulence intensities increase as the local Reynolds number $Re_r=\Omega r^2/\nu$ increases. Thus, for a given set of parameters ($Re_\Omega$, $G$), the successive transitions between regimes II and IV and then between regimes IV and III can be obtained [HAD08]. All the possible transition scenarii have been proposed by Cooper and Reshotko [COO75].

The diagram in Fig. 13 has been established by Daily and Nece [DAI60] from torque measurements. They obtained the following empirical correlations for the moment coefficient in each regime I, II, III or IV:

$$C_{M,I} = \pi \, G^{-1} \, \text{Re}_\Omega^{-1} \tag{38}$$

$$C_{M,II} = 1.85 \, G^{1/10} \, \text{Re}_\Omega^{-1/2} \tag{39}$$

$$C_{M,III} = 0.04 \, G^{-0.167} \, \text{Re}_\Omega^{-1/4} \tag{40}$$

$$C_{M,IV} = 0.051 \, G^{1/10} \, \text{Re}_\Omega^{-0.2} \tag{41}$$

The empirical correlation for regime I fully matches with the theoretical one of Owen and Rogers [OWE89]. Note that for regime II and in contrast with regime I, the moment coefficient increases then when *G* increases. Soo [SOO58] provides a theoretical correlation for the moment coefficient in the turbulent regime with merged boundary layers (regime III) with a dependence on *G* with a power -1/4 and a constant equal to 0.0308.

From the moment coefficient correlations, we can deduce for example, the aspect ratio for which the transition between regimes III and IV is observed:

$$G = 0.2112 \, \text{Re}_\Omega^{-3/16} \tag{42}$$

Owen [OWE84] proposed two empirical laws given the minimum and maximum values for which the boundary layers are merged and separated, respectively, in the whole cavity:

$$G_{min} = 0.23 \, Re_\Omega^{-0.2} \tag{43}$$

$$G_{max} = 1.05 \, Re_\Omega^{-0.2} \tag{44}$$

Daily and Nece [DAI60] and then Kreith [KRE68] proposed also different values of the critical Reynolds number above which the flow is turbulent in the entire cavity:

$$\text{Re}_\Omega > \left(\frac{\pi}{0.036}\right)^{4/3} G^{-10/9}, \quad \text{for } G < 0.0111$$
$$\text{Re}_\Omega > 6.9739 \times 10^6 G^{16/15}, \quad \text{for } 0.0111 < G < 0.0233$$
$$\text{Re}_\Omega > 1.266 \times 10^5, \quad \text{for } G > 0.0233$$

### 2.4.2 Instability patterns (regimes I and II)

During the transition process to turbulence in an enclosed rotor-stator cavity, a large variety of patterns can be observed according to the combination ($Re_\Omega$, *G*) as mapped by Schouveiler *et al.* [SCHOU01] in Fig. 17. Three different regimes have been observed by these authors: regimes I and II already introduced and a regime with mixed boundary layers. This third regime is characterized by merged boundary layers close to the rotation axis and unmerged ones at the periphery, and represents a very narrow band in the diagram in terms of aspect ratio: $0.0179 \leq G \leq 0.0714$.

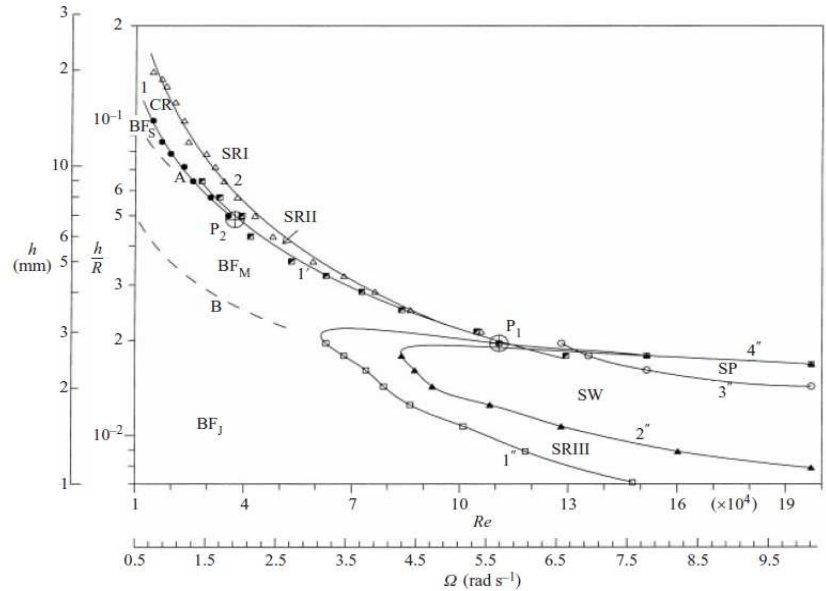

Figure 17: Transition diagram of the flow between a rotating and a stationary disk enclosed by a stationary sidewall. Curves A and B separate the mixed basic flow (BFM) from the basic flows with separated boundary layers (BFS) and with joined boundary layers (BFJ) respectively. Curves 1 and 2 are the thresholds for the circular rolls (CR) and the spiral rolls (SRI) respectively. Curve 1' is the spiral roll (SRII) threshold. Curves 1", 2" and 3" are the thresholds for the spiral rolls (SRIII), the solitary waves (SW) and the spots (SP) respectively. Curve 4" is the threshold for the simultaneous disappearing of the spiral rolls (SRIII) and of the solitary waves (SW). These thresholds have been determined by increasing the Reynolds number $Re_\Omega$, hysteresis cycles have been demonstrated for the thresholds 2" and 4", after Schouveiler *et al.* [SCHOU01].

In regime II ($0.0714 \leq G \leq 0.1429$), the stability of the flow is closely connected to that of the single disk. Experimental, numerical, and theoretical results display similar structures related to the two basic types of linear instability, referred to as Type I (or B) and Type II (or A), that can arise for different Rossby numbers. Type I is an inviscid instability related to an inflexion point in the profile of the radial velocity, whereas Type II instability is more specific to the rotating disk boundary layer because it is related to the combined effects of Coriolis and viscous forces and occurs at a lower critical Reynolds number than Type I instability [LIL66]. The presence of the hub and the shroud may indeed alter the global stability of the flow by introducing disturbances into the disk boundary layers. Moreover, the core-swirl ratio $\beta$, which can be seen also as a local Rossby number, is no longer constant and may vary with the radius [GAU99, RAND97, SER01], which affects the global stability properties of the boundary layers. The parameter $\beta$ also varies with the aspect ratio of the cavity as shown experimentally by Daily and Nece [DAI60]. They found $\beta$=0.44 and 0.36 for $G$=0.102 and 0.217 respectively at $Re_\Omega$=4.2×10$^4$, which is to be compared with the value $\beta$=0.313 obtained by Dijkstra and Van Heijst [DIJ83] for infinite disks.

The stability of Batchelor flows with unmerged boundary layers (regime II) is primarily governed by the disk boundary layers. The Bödewadt layer along the stator is besides much more instable than the Ekman layer along the rotor. Thus, all the works done by the IRPHE's team (SCHOU01, CRO02, CRO05, PON9b] have been performed using flow visualizations from above the stationary disk. The primary destabilization in the Batchelor regime occurs in the stator boundary layer in the form of axisymmetric circular rolls (CR) propagating towards the rotation axis and recognized as a Type II instability [SER04b], which was expected from linear stability theory for a single disk. For steady conditions in rotor-stator cavities, such vortices have been reported experimentally by Gauthier *et al.* [GAU99] and numerically using fully three-dimensional computations by Serre *et al.* [SER01] even for annular cavities. The instability is linearly convective, leading to a high sensitivity to external controlled or uncontrolled forcing [GAU99], which has been confirmed numerically by Poncet *et al.* [PON09b]. For increasing rotation rate, the axisymmetry of the flow is broken, and spiral waves, denoted SRI, develop in the Bödewadt layer with positive angles ranging from approximately 10° to 25° [SER01, SCHOU01]. Experimental [CRO05] and numerical results [PON9b] show that this is a supercritical Hopf bifurcation. At the lowest rotation rates, the spiral wave pattern can coexist with the previous circular waves that are noise sustained: spirals appear around the periphery and circles are observed at small radii. According to the linear stability analysis of the Batchelor flow by Serre *et al.* [SER04b], this spiral mode of positive angle is a Type I instability. Both experimental observations and numerical solutions typically exhibit spiral patterns with an azimuthal wave number in the range [16–30], depending on the value of the aspect ratio and rotation rate. Schouveiler *et al.* [SCHOU98] showed that the wave-number selection process of this secondary instability can be regarded as resulting from an Eckhaus instability that selects the number of spiral arms. The influence of a central hub on the characteristic of these two bifurcations is weak as shown by [PON09b].

The rotor layer is less well documented mainly because measurements are technically more difficult on that side. Moreover, instabilities occur at higher rotation rates, which make computations more computationally expensive due to resolution requirements. For annular cavities, Serre *et al*. [SER01] showed that the instabilities developed in the Bödewadt layer are convected along the hub and induce disturbances in the rotor layer, triggering axisymmetric and then three-dimensional convective modes characterized by a negative angle $\varepsilon$ with the tangential direction (-7.5 ≤ $\varepsilon$ ≤ −20**°)**.

The evolution to higher bifurcations in the rotor-stator cavity has received much less attention. To our knowledge, Cros *et al.*'s [CRO05] experiment is the only one that provides some insight into the evolution and further transitions of the stator boundary layer for Reynolds numbers up to around 74000. The transition to turbulence seems to be governed by nonlinear interactions of the circular and spiral modes leading to a kind of wave turbulence. Recent numerical studies provide some insight into transition in the rotor boundary layer. At high local Reynolds numbers ($Re_r$ > 400), the rotor layer becomes absolutely unstable, and Type I instability with spiral arms of positive inclination is observed. Large eddy simulations from Séverac *et al.* [SEV07a] confirm the presence of spiral arms in the rotor layer at $Re_r$ ≈ 400 with all the characteristics of the absolute Type I instability and, just upstream, a turbulent region. The different scenarios for the route to turbulence in such systems are discussed in Launder et *al.* [LAU10].

For the transition to turbulence in regime I with merged boundary layers ($G$ ≤ 0.0179), the single instability pattern reported before turbulence by Schouveiler *et al.* [SCHOU01] and Cros and Le Gal [CRO02] using flow visualizations and image processing is related to weakly negative spirals, denoted SRIII, that occur at the periphery of the cavity at large radius. The most interesting feature observed by Cros and Le Gal [CRO02] is that transition to turbulence occurs via defect turbulence in the network of negative SRIII spirals. Increasing further the rotation rate leads to the appearance of turbulent domains within a laminar background. These instabilities, in the form of turbulent spirals (SW) are periodically located around the disk and similar to the ones observed in plane Couette flows, Taylor-Couette flows or in the Ekman-Couette problem. By increasing progressively the rotation rate, turbulent spots (SP) appear and finally invade the whole cavity. Le Gal et *al.* [LEG07] showed by flow visualizations that these spots have a horse-shoe vertical structure and are generated in the area close to the gap between the rotating disk and the shroud. Transition mechanisms along the rotor are difficult to identify due to the speed at which turbulence ensues and due to the experimental techniques used.

### 2.4.3 Turbulent flow regimes (III and IV)

The flow in regimes II or IV exhibits a Batchelor-like structure with a rotating inviscid core separated by the Ekman layer on the rotor and the Bödewadt layer on the stator. The fluid flows outward along the rotor with a mean tangential velocity ranging from $\Omega r$ to $\beta \Omega r$. As there is neither radial nor axial flow in the core region, by conservation of mass, a radial inflow is created along the stator with a tangential velocity ranging from 0 on the disk and $\beta \Omega r$. One main difference between regimes II and IV stands in the value of the core-swirl ratio $\beta$. In regime IV, this coefficient is constant with the local radius $r$, as far as the two boundary layers are turbulent in the whole cavity. In regime II, it strongly varies with $r$. Stépanoff [STEP32] was the first to determine the core-swirl ratio $\beta$ of the turbulent flow in a rotor–stator cavity. He suggested a value of $\beta$ equal to 0.5 and independent of the radial location. Then Schultz-Grunow [SCHU35] proposed a theoretical value of $\beta$=0.512 and measured an experimental value equal to 0.357. He interpreted this discrepancy as due to the existence of a shear stress in the small radial gap between the rotating disk and the fixed cylindrical endwall. Using velocity and pressure measurements, Poncet *et al.* [PON05a] obtained $\beta$=0.438, which coincides with the asymptotic value given by their analytical model without throughflow. This coefficient appears to be independent of the interdisk spacing (the flow remaining in regime IV with unmerged boundary layers) and of the local radius but sensitive to the core-swirl ratio at the periphery of the cavity (prerotation level). For comparisons with available values of $\beta$ in shrouded or unshrouded systems, the reader can refer to the works of Randriamampianina *et al.* [RAND97] and Poncet *et al.* [PON08].

One other main difference between regimes II and IV is the thicknesses of the boundary layers. They behave globally like the thickness of the boundary layer over a single infinite rotating disk, $\delta = \sqrt{\nu/\Omega}$. Thus, both boundary layers get thinner for increasing rotation rates but not in the same way. The boundary layer thickness of the rotating disk is thinner than that of the stator. Both thicknesses decrease when one approaches the axis of the cavity for low radii *r*, as confirmed by the experimental laws of Daily and Nece [DAI60]: $\delta_E = r/\text{Re}_r^{1/5} f(G)$ for the Ekman layer and $\delta_B = 1.7 r/\text{Re}_r^{1/5} f(G)$ for the Bödewadt layer. The boundary layers are thus much thicker in enclosed systems in comparison with the boundary layer over a single rotating disk.

The experiment of Itoh *et al.* [ITO92] provided a great contribution to the understanding of the turbulent flow in a shrouded rotor-stator system in regime IV. The authors measured the mean flow and all the Reynolds stress components for *G*=0.08 and brought out the existence of a relaminarized region even for high rotation rates. Cheah *et al.* [CHE94] performed detailed measurements of mean velocity and turbulence intensities inside a rotor-stator system for *G*=0.127. These two experiments have been used as a reliable experimental database to evaluate the performances of turbulence models as in Elena and Schiestel [ELE95] or in Iacovides *et al.* [IAC96] for examples. More recently, Poncet *et al.* [PON05b] compared the predictions of the RSM model of Elena and Schiestel [ES96] with velocity and pressure measurements in the case of a closed rotor-stator system of aspect ratio *G*=0.036 and rotational Reynolds numbers up to $Re_\Omega=4.15\times10^6$. From all these works, it appears that turbulence is mainly concentrated within the boundary layers with very low turbulence levels in the inviscid core. The Ekman layer is besides more turbulent than the Bödewadt layer. The Bödewadt layer gets turbulent at lower radii than the Ekman layer. Itoh *et al.* [ITO92] attributed the difference to a greater stability of the radially outward flow in the Ekman layer than the return flow in the Bödewadt layer. For Cheah *et al.* [CHE94], it is due to the effects of the convective transport of turbulence. The fluid in the Bödewadt layer comes from the periphery of the cavity where the highest turbulence levels are observed due to the highest values reached by the local Reynolds number $Re_r=\Omega r^2/\nu$. Then, it flows radially inward along the stator with decreasing intensity levels. On the contrary, the fluid in the Ekman layer comes from the rotation axis, where the lowest values of $Re_r$ are obtained. For sufficiently high rotation rates, both boundary layers are turbulent in the whole cavity and the rotational Reynolds number $Re_\Omega$ has no significant influence on the turbulence intensities. Reynolds stress tensor components are strongly anisotropic near the walls and the anisotropy then diminishes in the core region.

Regime III is also less documented than regime IV. The main experimental investigation is the one of Daily *et al.* [DAI64], who performed mean velocity, torque, and pressure measurements, for $G$=0.0138 and two Reynolds numbers $Re_\Omega$ =2.95–6.9×$10^5$. For this set of parameters, they clearly obtained merged boundary layers. Phadke and Owen [PHA88] studied, using flow visualizations and pressure measurements, the effect of seven shroud geometries on the ingress of external fluid into the rotor-stator cavity for a large range of flow control parameters including the torsional Couette flow regime. More recently, Andersson and Lygren [AND06] performed some LES of enclosed rotor-stator flows for both the wide and narrow gap cases. They obtained a turbulent torsional Couette flow for $Re_e$=160, which corresponds to three sets of parameters: ($G$=0.02, $Re_\Omega$=4×$10^5$), ($G$=0.01, $Re_\Omega$=1.6×$10^6$) and ($G$=0.01265, $Re_\Omega$=$10^6$). The first set has been studied experimentally by Itoh [ITO95]. For this value of $Re_e$, the dimensionless mean tangential velocity exhibits an S-shaped profile, with a value consistently below 0.5 midway between the disks. In the narrow-gap cases, $u_\theta/(\Omega r)$ is 0.43, 0.46 and 0.47 in the three different cases and thereby tended towards the plane Couette flow value 0.5 with increasing values of $Re_\Omega$. The profiles of mean radial velocity show that the crossflow is substantially weaker than in the laminar case and reduces with increasing $Re_\Omega$. Haddadi and Poncet [HAD08] proposed extensive computations using the RSM of [ES96] for a wide range of the flow control parameters (1.8×$10^5$ ≤ $Re_\Omega$ ≤ $10^7$, 0.02≤ $G$ ≤ 0.05) mainly covering regime III. The axial profiles of the mean tangential velocity case remain very similar to the ones obtained for turbulent Batchelor flows (regime IV). On the contrary, the profiles of the radial velocity are quite different. It is always linear in the main part of the flow as in classical plane Couette flows. Contrary to turbulent Batchelor flows, where turbulence is concentrated in the boundary layers, turbulence in torsional Couette flows is distributed along the axial direction. Moreover, the levels of the normal stresses are quite comparable, which is not the case for Batchelor flows, where the axial normal component of the Reynolds stress tensor is generally negligible compared to the other normal components [PON05a].

### 2.4.4 Heat transfer

Nikitenko [NIK63] conducted experiments with an air-filled enclosed rotor–stator system where both disks were isothermal for a wide range of the aspect ratio 0.018≤$G$≤0.085 and of the Reynolds number $Re_\Omega$≤10$^6$. He found no effect of the aspect ratio and correlated his results for the rotor side according to $Nu_r = 0.675\left((r/R)^2 Re_\Omega\right)^{1/2}$ in the laminar regime and to $Nu_r = 0.0217\left((r/R)^2 Re_\Omega\right)^{4/5}$ in the turbulent one, with $Nu_r$ the local Nusselt number. Shchukin and Olimpiev [SHC75] measured the mean Nusselt number in a closed rotor-stator cavity such as $G$=0.0646. For a radial temperature distribution on the rotor varying as $r^{0.25}$, their results for the turbulent regime were correlated by $\overline{Nu_R} = 0.0168\, Re_\Omega^{0.8}$, which is 11% higher than the free disk value. Many later works will show that both the local and averaged Nusselt numbers depend on the aspect ratio of the system.

In regime I corresponding to a laminar flow with merged boundary layers, Owen and Rogers [OWE89] proposed a correlation for the local Nusselt number on the rotating disk, depending on the aspect ratio of the cavity and not on the rotational speed:

$$Nu_{r,I} = \frac{1}{G}\frac{r}{R} \Rightarrow \overline{Nu_I} = \frac{1}{G} \tag{45}$$

The local Nusselt number proportionally decreases with the increase in $G$ for regime I. When the Reynolds number increases, the flow becomes turbulent (regime III) and heat transfer depend on the rotational velocity and $G$ such as:

$$Nu_{r,III} = 0.01176\left(\frac{r}{R}\right)^{7/4} G^{-1/4} Re_\Omega^{3/4} \Rightarrow \overline{Nu_{III}} = \frac{0.0308}{\pi} G^{-1/4} Re_\Omega^{3/4} \tag{46}$$

Dorfman [DOR63] proposed correlations for the local Nusselt numbers on the rotor for regimes II and IV with unmerged boundary layers:

$$Nu_{r,II} = 0.922\, Re_r^{0.5} \tag{47}$$

$$Nu_{r,IV} = 0.0251\, Re_r^{0.8} \tag{48}$$

In his experiment, the aspect ratio $G$ was fixed, that is why it does not appear in these correlations. Owen and Rogers [OWE89] observed that the global heat transfer over the rotating disk is minimum when the flow regime is in transition between the low spacing regimes and the high spacing regimes. When *G* increases, the Couette-type flow disappears and a rotating core of fluid appears which strongly decreases the shear stresses in the air-gap. When *G* increases again, the rotating core of fluid disappears, which is the transition between Batchelor and Stewartson type flow, increasing the shear stresses so the convective heat transfer. Daily and Nece [DAI60] proposed correlations for the mean Nusselt numbers on the rotor:

$$\overline{Nu_{II}} = \frac{2}{\pi}\left(\frac{G}{2}\right)^{0.1} Re_\Omega^{0.5} \tag{49}$$

$$\overline{Nu_{IV}} = \frac{0.0545}{\pi}\left(\frac{G}{2}\right)^{0.1} Re_\Omega^{0.8} \tag{50}$$

Pellé and Harmand [PEL07] performed extensive measurements of the heat transfer coefficients along the rotating disk in a rotor–stator system with an open end air-gap for rotational Reynolds numbers $Re_\Omega$ in the range [$1.29 \times 10^5$ - $6.45 \times 10^5$] and aspect ratios *G* ranging from 0.01 and 0.16. They correlated their results for the local Nusselt numbers in regimes I, II and IV as follows:

$$Nu_{r,I} = 70\left(1 + e^{-140G}\right) Re_\Omega^{-0.456} Re_r^{0.478} \tag{51}$$

$$Nu_{r,II} = 0.463\left(1 - e^{-40G}\right)\left(1 - e^{-120000 Re_\Omega}\right) Re_r^{0.746} \tag{52}$$

$$Nu_{r,VI} = 0.035\left(1 - e^{-40G}\right)\left(1 - e^{-420000 Re_\Omega}\right) Re_r^{0.746} \tag{53}$$

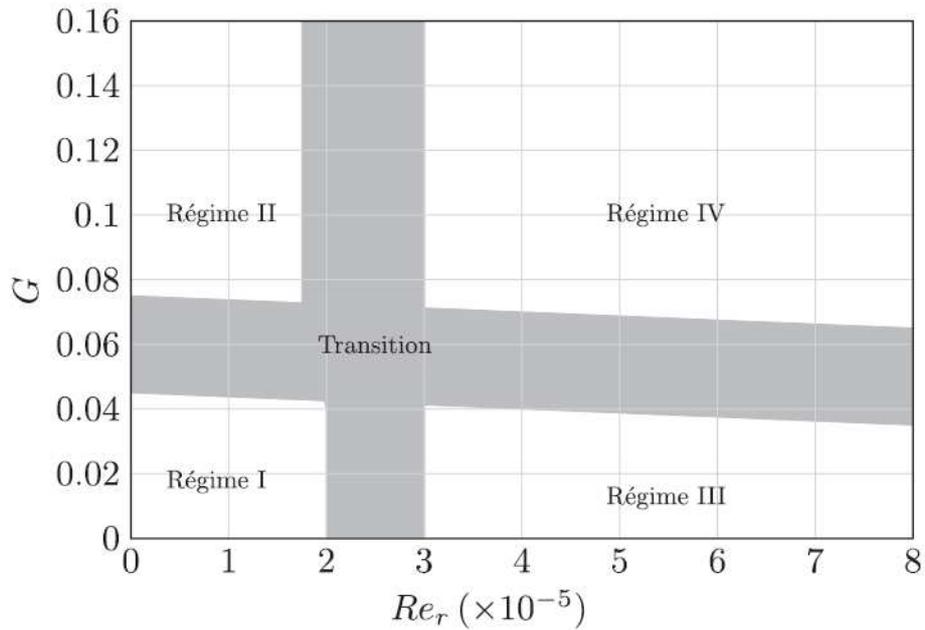

Figure 18: The four heat transfer regimes depending on G and Re as established by Pellé and Harmand [PEL07].

From a general point of view, their results showed that none of the previous correlations obtained either by Owen and Rogers [OWE89] for regimes I and III or by Dorfman [DOR63] for regimes II and IV, provide fully satisfactory results for the whole disk surface, whatever the set of parameters ($Re_\Omega$, G). For G=0.01, the local Nusselt numbers remain very low and do not vary so much with the rotation rate. Thus, they are well approximated by the correlation relative to regime I [OWE89], which is not the case for the other values of *G*. For the highest interdisk spacings, the local Nusselt numbers are found to be very similar to those obtained in the case of a single disk for all considered values of the rotational Reynolds number. Pellé and Harmand [PEL07] proposed also a method to calculate the averaged Nusselt numbers over the whole disk taking into account the possible transition between two regimes for a given set of parameters. They provided then correlations for the mean Nusselt numbers for each regime together with a schematic representation of the four heat transfer regimes, as Daily and Nece [DAI60] did it before for a closed cavity under isothermal conditions (see Fig. 18).

There is much less results concerning the heat transfer along the stationary disk. We can cite the work of Yuan *et al.* [YUA03], who performed measurements using the liquid crystal technique to determine the turbulent heat transfer on the stationary disk in a rotor-stator cavity opened at the periphery for $1.42\times10^5 \leq Re_\Omega \leq 3.33\times10^5$. They provided also axisymmetric computations using a RNG k-ε model available within Fluent, assuming no thermal effect on the hydrodynamic flow. They focused their attention to the flow and thermal fields at the periphery of the stator. Like open-end rotor-stator flows, the mean velocity maps exhibit a Stewartson-like structure, confirming the results of Brady and Durlofsky [BRA87]. It has an important role on the heat transfer process. For each value of $Re_\Omega$, their results highlighted the existence of an optimum interdisk spacing at which the averaged Nusselt number on the stator was maximum. The optimum spacing is found to decrease when the Reynolds number increases. Howey et al. [HOW10] developed a new experimental method for measuring stator heat transfer in a rotor–stator system partially blocked at the periphery for $G$=0.0106, 0.0212 and 0.0297 and $3.7\times10^4 \leq Re_\Omega \leq 5.6\times10^5$. Transition at the stator was observed to occur at $Re_\Omega > 3\times10^5$ for all aspect ratios. An Increase of the Nusselt numbers at the periphery was observed for all sets of parameters due to the ingress of ambient fluid along the stator because of the rotor pumping effect. Their results highlighted so the importance of peripheral fluid inflow at the stator in determining the stator heat transfer at outer radii.

Other effects like the prerotation level or the presence of a large central opening in the stator on the heat transfer have also been considered. Dibelius and Heinen [DIB90] investigated experimentally the heat transfer along the rotor in a quasi-isothermal-rotor / cooled-stator cavity for $Re_\Omega$=2×10$^6$ and three aspect ratios $G$=0.0125, 0.0625 and 0.1375. They obtained two main results: the heat transfer increases as $G$ decreases in contradiction to the application of the Reynolds analogy and most of all, the heat transfer are reduced in the presence of prerotation at the cavity inlet. Beretta and Malfa [BER03] studied the effect of a central opening on the stator side on the heat transfer in a rotor-stator system. They performed computations using a low Reynolds number k-ε model within Fluent for adiabatic-rotor / isothermal-stator conditions and compared them to an in-house semi-empirical model developed to evaluate temperature and heat fluxes and based on mass and angular momentum balances and the Reynolds analogy. A reasonable agreement was found for both geometries: closed and opened housings. They showed in particular that the radial distribution of temperature difference between rotor and stator disks is parabolic and that, under conditions of validity of the Reynolds analogy, it is independent of the presence or not of a radial outflow. Boutarfa and Harmand [BOU03] investigated also the influence of a central opening in the stator on the flow structure and the local convective heat transfer in a heated-rotor / isothermal-stator enclosure. They used PIV and infrared thermography respectively for $5.87 \times 10^4 \leq Re_\Omega \leq 1.76 \times 10^5$ and $0.01 \leq G \leq 0.17$. The originality of the system, previously considered by Harmand *et al.* [HAR00] for $2 \times 10^4 \leq Re_\Omega \leq 1.47 \times 10^6$ and $G = 0.01$, consisted in a large central opening in the stator. By comparisons with the previous results of Daily and Nece [DAI60], they highlighted the influence of the central opening, which increases the heat transfer coefficient, whatever the values of $Re_\Omega$ and $G$. For a narrow gap cavity $G=0.01$, the flow exhibits a torsional Couette-like structure with an outward flow in the gap, which enhances the heat transfer by comparison with large aspect ratio cavities. For $G=0.02$, the flow switches to a Batchelor structure, which induces a decrease of the tangential shear stresses on the rotor and consequently of the Nusselt number. When $G$ increases further, the Nusselt number on the rotor is close to that obtained for a single disk. They proposed correlations for the local and averaged Nusselt numbers along the rotor as a function of the rotational Reynolds number and the aspect ratio.

### 2.5. Rotor Stator with jet

In this section, a confined round jet impinges onto a rotating disk (Figure 19). The interaction of rotation and impingement creates a complex and powerful flow capable of improving heat transfer processes considerably. The flow is confined by a stationary disk located at an axial distance $e$ of the rotating disk. The incoming fluid is discharged from the center of the stator through an orifice of diameter $D$ at a volumetric flow rate $Q$. As the fluid is the same as the fluid within the cavity, it is called submerged jet. The rotor-stator cavity may be partially enclosed at the periphery by a stationary cylinder (shrouded cavity) or opened to atmosphere (unshrouded cavity).

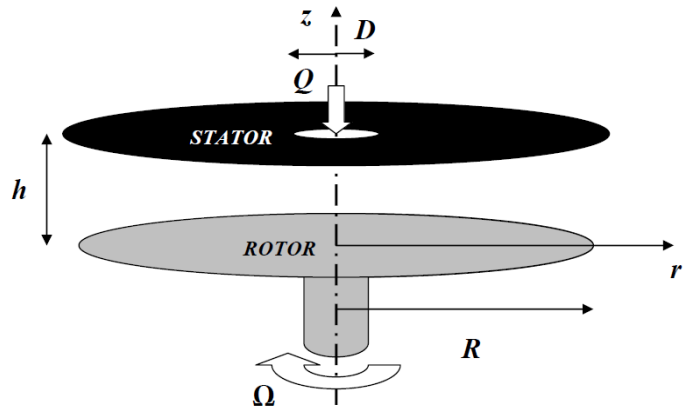

Figure 19: Schematic diagram of a cylindrical rotor-stator cavity with an impinging round jet.

### 2.5.1 Hydrodynamic field

*Round impinging jet without rotation*

The configuration without rotation of the bottom disk is often used as a reference test case to validate experimental or numerical methods. For an exhaustive review about impinging jet flow in so-called stator-stator cavities, the reader may refer to the recent works of [CHEN02, HSI06] for examples.

Confined submerged liquid jets in either axisymmetric or planar share the common feature of a small stagnation zone at the impingement surface whose size is of the order of the nozzle diameter *D* or slot dimension, with the subsequent formation of a wall jet region.

Cheng *et al.* [CHEN02] established the transition diagram in a ($Re_j$, $Ra$) plane and provided extensive flow visualizations of the flow patterns due to the combined effects of buoyancy and jet impingement onto a heated disk in a closed cylindrical system. They investigated in details the influence of the flowrate, the temperature difference and the pressure inside the cavity. The typical flow pattern appears in the form of two axisymmetric circular vortex rolls, one around the air jet along with another roll near the side wall of the enclosure. Both rolls have deformed cross-sections (Figure 20). The rolls are generated by the reflection of the jet from the impingement surface and by the deflection of the wall boundary layer flow along the surface by the upward buoyancy due to the heated plane. A rise in the flowrate causes the inner vortex roll to become substantially larger and the outer vortex roll becomes correspondingly smaller. Increasing the temperature difference between impingement surface and the air jet has an opposite effect.

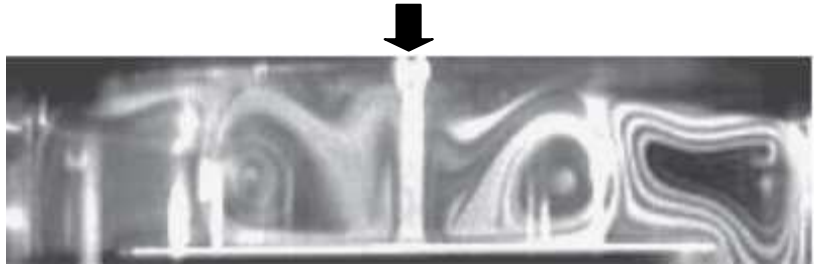

Figure 20: Flow visualizations of the impinging jet flow for $Re_j$=520.1, $P$=450 Torr and $\Delta T$=15 K after Cheng *et al.* [CHE02].

Hsieh and Lin [HSI05] performed flow visualizations and temperature measurements in order to investigate the influence of the jet to disk separation distance on the flow patterns. A small tertiary inertia-driven circular roll is identified in the confined impinging jet for high jet Reynolds numbers. Beyond a critical value of $Re_j$, the high jet inertia causes the vortex flow to become time dependent. A reduction in the interdisk spacing results in a significant decrease in the Rayleigh number and can substantially reduce the size and strength of the buoyancy-driven roll. The time-dependent flow behaviour has been considered in more details by Hsieh *et al.* [HSI06] for $Re_j$ ≤1623, Ra ≤ 63420 and $e$=10, 20 and 30 mm. At sufficiently high $Re_j$, the inertia-driven tertiary and quaternary rolls can be induced aside from the primary and secondary rolls. At an even higher $Re_j$ the vortex flow becomes unstable due to the inertia driven flow instability. For $e$ = 20 mm, the flow is also subjected to a buoyancy-driven instability for this range of parameters (Figure 19). They finally proposed flow regime maps delimiting the temporal state of the flow together with correlating equations for the boundaries separating various flow regimes.

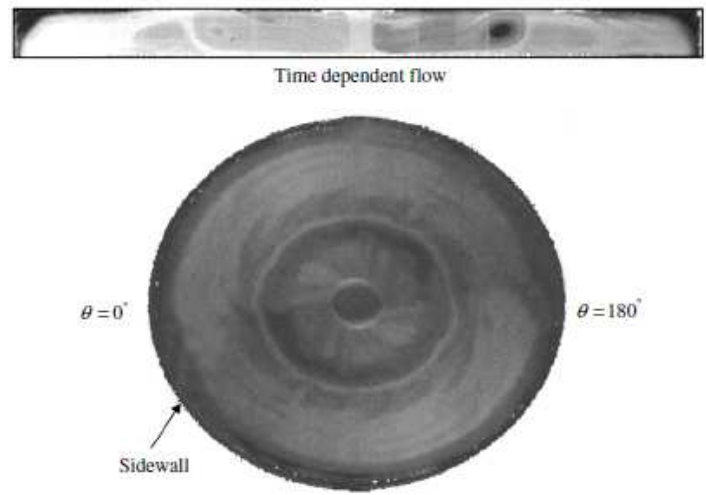

Figure 21: Time-dependent impinging jet flow for $Re_j$=829, e=20 mm, $t$=1.45 s in the isothermal case after Hsieh *et al.* [HSI06].

***Transition between the Batchelor and the Stewartson flow structures***

When an impinging jet is superimposed on the base rotor-stator flow, the same four flow regimes already discussed in the previous section have been observed with similar but different boundaries. In particular, a superposed radial outflow can cause a premature transition to turbulence [KRE63, OWE89]. The second difficulty, which arises when dealing with opened cavities, is the possibility for the external fluid from the surroundings to be pulled into the system (ingress), when the superposed flowrate is relatively weak.

Soo [SOO58] obtained approximate solutions for regime I and displayed the different flow patterns according to only one parameter $\Phi(r/R)$:

$$\phi = \frac{G\ Cw}{2\pi\ Re_\theta^2}\left(\frac{R}{r}\right)^2 \quad (54)$$

with $\Phi$<0 for inward superposed flow and $Cw = Q/vr$. For $\Phi$ =1/90, the flow is purely centrifugal, whereas some ingress is observed along the stator for $\Phi$ =1/180. Köhler and Müller [KOH71] compared velocity measurements with calculations based on a finite-difference method for G=0.01 and $Re_\Omega$<1200 with a reasonable agreement between the two approaches.

In regime III, Dorfman [DOR61] obtained approximate solutions using an integral method for the case of turbulent flow without an inviscid core. He provided interesting correlations for the moment coefficient for the two limiting cases without throughflow (*Q*=0) and for *Q*→∞. Haddadi and Poncet [HAD08] proposed RSM calculations for 0.02 ≤ *G* ≤ 0.05, 1.8×10$^5$ ≤ *Re$_\Omega$* ≤ 10$^7$ and *C$_w$* ≤ 10$^4$. In most cases, the flow belongs to the regime III: turbulent with merged boundary layers. Turbulence is mainly confined along the rotor and vanishes toward the stator. The maximum of the turbulence intensities is obtained at the inlet close to the rotation axis and turbulence levels slightly decrease when moving toward the periphery of the cavity. It is mainly composed of three components, whatever the values of the flowrate coefficient *Cw*.

In regime II with laminar and unmerged boundary layers, an inviscid rotating core may appear inside the cavity between the two boundary layers. The flow structure still depends on the Reynolds number *Re$_\Omega$* but also on a new parameter, the flow rate coefficient *C$_w$*, which are combined to form the flow parameter *κ*. For a laminar flow, this parameter is defined as:

$$\kappa_{lam} = C_w \operatorname{Re}_\Omega^{-0.5} \tag{55}$$

Soo [SOO58] has shown that the tangential velocity of the fluid rotating core is a function of the radial location. For radii larger than $(\kappa_{lam}/\pi)^{0.5}$ the rotating core does not appear and the flow near the rotor is similar to the one obtained for a single rotating disk. Soo [SOO58] has furthermore said that for $(\kappa_{lam}/\pi)^{0.5} < 1$, the flow exhibits a Batchelor-like profile and for $(\kappa_{lam}/\pi)^{0.5} > 1$, it switches to a Stewartson-like structure, with only one boundary layer on the rotating disk.

In regime IV (turbulent flow with unmerged boundary layers), the structure of the turbulent flow is similar to the previous case. Some authors have defined a new flow parameter as being:

$$\kappa_{tur} = C_w \operatorname{Re}_\Omega^{-0.8} \tag{56}$$

In order to characterize the flow, Owen and Rogers [OWE89], like Soo [SOO58] for configuration II, have defined a parameter, such that $(\kappa_{turb}/0.219)^{5/13}$. For radii larger than $(\kappa_{turb}/0.219)^{5/13}$, the inviscid core region becomes negligible, and the flow is a Stewartson-type flow.

Both Batchelor and Stewartson flow structures have been obtained experimentally and numerically in only very few works. We can cite, among others, Elena and Schiestel [ELE93a], who proposed calculations using both a k-ε and an algebraic stress model (ASM) for the turbulent impinging jet in a rotor-stator cavity with a central hub. In their case, the jet impinges the rotor at a given radial distance of the rotation axis to model the experiments of Daily *et al.* [DAI64]. They obtained satisfactory results for $Re_\Omega$=2.9x10$^5$ and 6.9x10$^5$ and $C_w$=3530 and 7061. For $Re_\Omega$>6.9x10$^5$, the flow structure gets of Stewartson type with an outward flow whatever the axial position and a quasi zero tangential velocity outside the rotor layer. For $C_w$=3530, there is no Ekman layer on the rotor at $Re_\Omega$=10$^5$ and the radial velocity varies linearly in the core. When increasing the rotational Reynolds number, the Ekman layer develops and a small inward throughflow appears along the stator. When increasing further the rotation rate, the Batchelor structure is more evident and the core-swirl ratio tends to the limit 0.4 obtained by Poncet *et al.* [PON05a] for $Re_\Omega$=10$^7$. The same observations can be made at constant rotation rate by decreasing the flowrate coefficient.

The transition between the Batchelor and the Stewartson flow structures has been considered in details by Daily *et al.* [DAI64] and more recently by Poncet *et al.* [PON05b]. Daily *et al.* [DAI64] correlated their experimental results for the core-swirl ratio by:

$$\frac{\beta}{\beta_0} = \left(1 + 12.74 \frac{\kappa_{turb}}{(r/R)^{13/5}}\right)^{-1} \quad (57)$$

where $\beta_0$ is the core-swirl ratio in the case without throughflow. Poncet *et al.* [PON05b] established a new analytical law for the core-swirl ratio β for a turbulent rotor-stator in an annular enclosure submitted either to a centripetal or centrifugal injection:

$$\beta = 2 \times (5.9 \times Cq_r + 0.63)^{5/7} - 1 \quad (58)$$

where $Cq_r = Q \operatorname{Re}_r^{1/5}/(2\pi r^3 \Omega)$ is a local flowrate coefficient. This law is still valid for a weak centrifugal injection as long as the flow remains turbulent with unmerged boundary layers (regime IV). It has been validated against velocity and pressure measurements and also against the predictions of a RSM model [PON05a]. It has been extended to the laminar case by Poncet *et al.* [PON08]. For high values of the superimposed outflow, the flow structure switches to a Stewartson-like structure depending on the Rossby number $Ro = Q/(2\pi j R^2 \Omega)$, $j$ being the radial gap between the rotor and the shroud. For $G$=0.036, the critical value of Ro for the Batchelor / Stewartson transition in regime IV is given by [PON05b] as:

$$Ro = 0.0088 - 0.0998\, r/R + 0.3048\, (r/R)^2 - 0.4646\, (r/R)^3 \tag{59}$$

Very recently, Nguyen et *al.* [NGUYEN12] provided the first quantitative experimental database for the mean and turbulent flow fields in the case of an impinging jet in an open rotor-stator system (G = 0.02, e/D = 0.25) for $0.33\times10^5 \leq Re_\Omega \leq 5.32\times10^5$ and $17.2\times10^3 \leq Re_j \leq 43\times10^3$ by the means of PIV measurements. A recirculation flow region, which was centered at the impingement point and possessed high turbulence intensities, was observed. Local peaks in root-mean-square fluctuating velocity distributions appeared in the recirculation region and near the periphery, respectively with a weak influence of the rotational speed (Fig.22). By POD velocity decompositions, their results revealed the existence of coherent large-scale structures with an oval shape and a size comparable to the interdisk spacing. Moreover, they showed that they appear in all configurations with and without rotation of the bottom disk.

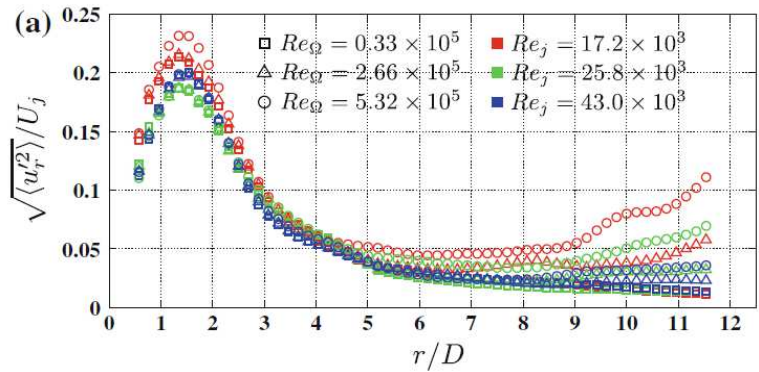

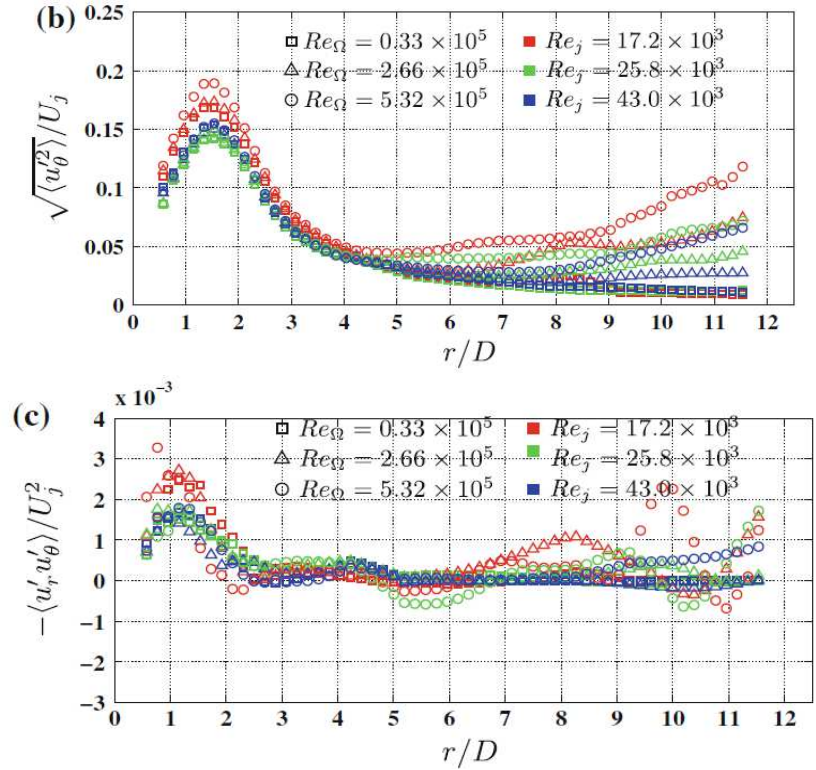

Figure 22: Non-dimensionalised r.m.s fluctuating velocities obtained by PIV measurements near the rotor for various rotational speeds and jet velocities after Nguyen et al. [NGUYEN12].

*Confinement effects*

If the rotor-stator cavity is unshrouded, the pumped flow from the rotating disk exits the cavity to the external surroundings and inflow (ingress) from the external environment may occur to supply the entrainment demands of the rotating flow. At the outer radius of the cavity there can therefore be both inflow on the stator and outflow on the rotor. The case of an unshrouded rotor-stator cavity with radial outflow has been studied for laminar merged boundary layers by Soo [SOO58], for laminar but unmerged boundary layers by Vaughan [VAU87], for turbulent merged boundary layers by Dorfman [DOR61], and for turbulent separate boundary layers by Daily *et al.* [DAI64].

The addition of a shroud at the periphery of a rotor-stator cavity normally acts to restrict the fluid exchange between the cavity and the external environment and so ingress. Bayley and Owen [BAY70] and then Haynes and Owen [HAY75] investigated the impact of a shroud attached at the stationary disk. For $Re_\Omega < 4.2 \times 10^6$, the shroud has the effect to reduce the moment on the rotor when no superposed flow is supplied; however, it has the opposite effect when a superposed outflow is supplied. This behaviour was attributed to the relative strengths of both increased angular momentum conservation due to the better sealing of the cavity and increased viscous shear as the flow exits the cavity due to the narrow gap. At very low values of the flowrates, the conservation of angular momentum in the cavity is dominant. As the flow is increased the shear in the rim region at the periphery becomes more significant. The moment on the rotor drops for large shroud to rotor gap. The overall gap between the rotor and stator disks has little impact on disk torque. As with the unshrouded case, the moment coefficient drops with increasing rotational Reynolds number [BAY69]. The presence of the shroud increases the pressure in the cavity above the ambient conditions. The pressure inside the cavity increases as the shroud clearance ratio and rotational Reynolds number decrease. For small gap ratios, an experimental study was undertaken by Bayley and Owen [BAY70] to determine the minimum superposed flow requirements to fully seal a cavity by measuring the pressure difference across the shroud. They assumed that the inflow was suppressed when the pressure difference between a location under the shroud and the external environment was zero. The superposed flow necessary to prevent ingestion $Cw_{min}$ was found to be strongly dependent on the shroud clearance $G_c$, independent of gap ratio and well predicted by the empirical relationship:

$$Cw_{\min} = 0.61 G_c \operatorname{Re}_\Omega \tag{60}$$

for $G_c=e_c/R=0.0033$ and $0.0067$ ($e_c$ the axial clearance between the shroud and the rotor), $G=0.06$, $0.12$ and $0.18$ and $4\times10^5 < Re_\Omega < 4.2\times10^6$. The results of Bayley and Owen [BAY70] for the moment coefficient were confirmed numerically by Gosman *et al.* [GOS76] using a k-ε model with reasonable agreement.

Bayley and Owen [BAY69] studied the effects of a radial outflow in a rotor-stator cavity by comparing experimental measurements of pressure, velocity and moment coefficients with values obtained from integral solutions of the boundary layer equations. Tests were performed at $Re_\Omega<3.8\times10^6$ for a wide range of aspect ratios $G$ from 0.008 to 0.03. As already discussed, the main effect of the outward flow created by the jet is the reduction of the core-swirl ratio, which can tend to zero at larger aspect ratios, and then a Stewartson-like structure occurs. In the same way, increasing the rotational Reynolds number increases the tangential velocity within the cavity. The main new result concerned the influence of the interdisk spacing $e$ on the flow structure. As $e$ is increased, the tangential velocity of the inviscid core region reduced and ultimately the core region disappeared as the effects of the stator reduced and the flow began to behave as if adjacent to a single free rotating disk. In a latter work, Owen *et al.* [OWE74] proposed the limiting aspect ratio *G* beyond which the moment coefficient on the rotor is unaffected by the presence of the stator, for an unshrouded system: $G = 1.05 \text{ Re}_\Omega^{-0.2}$ with *Re$_\Omega$* up to 4.1 millions, aspect ratios *G* in the range [0.005-0.604] and nozzle diameter to interdisk spacing ratio between 51 and 2.25. Bayley and Owen [BAY69] showed that narrower gaps resulted in higher moment coefficients. The pressure distribution predicted by momentum integral techniques agreed well with experimental data in the larger interdisk spacings but, as *e* was reduced, the effect of the outflow was overpredicted. Conversely, the prediction of moment coefficient improved for small clearances.

*Mass transfer data*

Kreith *et al.* [KRE63] performed mass transfer measurements for 0.12<*G*<0.6, $C_w \leq 1.51 \times 10^5$, $5 \times 10^3 \leq Re_\Omega \leq 10^5$. They proposed a correlation for the averaged Sherwood number:

$$\overline{Sh} = \frac{\overline{k}R}{Cd} = \left(\frac{G}{2}\right)^{0.55} \left(\frac{Cw}{2\pi G}\right)^{0.88 - \frac{12 Re_\Omega}{10^5}} \Gamma(Re_\Omega) \tag{61}$$

where *h$_{mav}$* is the averaged mass transfer coefficient, *D$_v$* the diffusion coefficient for naphthalene in air and *Γ* a fourth order polynomial law. Sara *et al.* [SAR08] used the electrochemical limiting diffusion current technique to measure the mass transfer from an impinging jet to a rotating disk for rotational Reynolds numbers in the range *Re$_\Omega$*=[3.4×10$^4$– 1.2×10$^5$], jet Reynolds number *Re$_j$* between 1.7×10$^4$ and 5.3×10$^4$ and non-dimensional jet-to-disk spacing *e/D*=[2–8]. They found that the jet impingement resulted in a substantial enhancement in the local and averaged mass transfer coefficients compared to the case without jet. The effect of jet impingement is dominant compared to the one of the rotation rate in this range of flow parameters. Mass transfer rates increase with increasing e/D, have a maximum value for *e/D* between 5 and 6, and then decrease with increasing *e/D* depending on both *Re$_\Omega$* and *Re$_j$* (Figure 23).

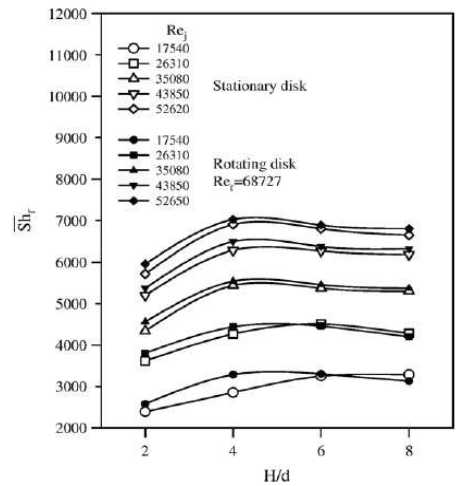

Figure 23: Average Sherwood number as function of non-dimensional jet spacing for stationary and for rotating disk after Sara *et al.* [SAR08].

### 2.5.2 Heat transfer

In the literature, a huge amount of heat transfer data is available in the case of an impinging jet onto circular disks. These data are often restricted to either cooling of a stationary disk by jet impingement or by pure rotation. The effect of the combination of rotation and jet impingement has been considered in only few papers as we will see in the following. Garimella and Nenaydykh [GAR96] and Li *et al.* [LI97] considered a confining top plate for a submerged liquid jet at different flowrates without any rotation. They determined experimentally the effects of the nozzle geometry on the local heat transfer coefficients from a small heat source to a normally impinging round submerged confined jet at different nozzle–to–plate spacing and jet Reynolds numbers. Rahman *et al.* [RAH02] numerically evaluated the conjugate heat transfer of a confined jet impingement over a stationary disk. Ichimiya and Yamada [ICHI03] presented the heat transfer and fluid flow characteristics of a single circular laminar impinging jet, including buoyancy effect in a narrow cavity. They identified the presence of forced, mixed, and natural convection modes of heat transfer as the flow moved downstream in the radial direction. The reader can refer to the work of Webb and Ma [WEB95], who presented a comprehensive review of studies on jet impingement heat transfer up to 1995. They concluded that heat transfer in submerged jets is more sensitive to nozzle–to–disk spacing than in free jets, especially when the heat transfer surface is beyond the potential core of the jet. Impinging jet in rotating cavities may be used also as an effective way to heat fluids. For example, Ellwood and Korchinsky [ELW00] compared measurements and theoretical predictions of viscous dissipation rates for a shrouded rotor-stator cavity of aspect ratio $G$=0.0093 with an impinging jet. They considered two viscous fluids, an automotive anti-freeze, ethylene glycol and a diluted glycerol to quantify the viscous heating due to the rotation of the disk.

Only few groups of researchers have concentrated their efforts to investigate the problem of a jet impinging onto a rotating disk with confinement effects due to both a stationary disk facing the rotor and to an external stationary cylinder enclosing the cavity. Yu *et al.* [YU73] performed flow visualizations and heat transfer measurements along the rotating disk in a shrouded rotor-stator enclosure for $G$=4/9, 8/9 and 2, $Re_\Omega$=1.8×10$^5$, 5.4×10$^5$ and 1.1×10$^6$ and $C_w$=23560, 37700 and 73830. For some combinations of ($G$, $Re_\Omega$, $C_w$), they observed strong asymmetries in the flow patterns. They found that the Nusselt number increase by increasing either the rotation rate or the coolant flowrate or by decreasing the interdisk spacing. In the laminar regime, the heat transfer coefficient is insensitive to the flowrate coefficient. Haynes and Owen [HAY75] extended their work to the heat transfer problem for $G$=0.055, 4×10$^5$ ≤ $Re_\Omega$ ≤ 4×10$^6$, 3.1×10$^4$ ≤ $C_w$ ≤ 7.5×10$^4$ and two dimensionless axial clearances $G_c$=0.0033 and 0.0067 between the shroud and the rotor. The averaged Nusselt number on the rotor increases for decreasing values of $G_c$. For large values of the rotational Reynolds number, the results tend towards the free disk case, and for small values, they are quite independent of the rotation rate. Sparrow and Goldstein [SPA76] performed well controlled heat transfer measurements in a shrouded rotor–stator system with an axial outward throughflow for the same parameters as in Yu *et al.* [YU73]. They showed in particular the strong influence of the coolant flowrate on the heat transfer coefficient along the shroud and the relative importance of rotation and aspect ratio of the cavity. For large flowrates, the heat transfers are independent of the presence or absence of rotation. They are found to increase with increasing values of the flowrate coefficient. For low rotation cases, the temperature field remains quite uniform within the cavity and large non uniformities appear at strong rotation rates. Finally, they showed by comparisons with their previous work [SPA75] that the position of the outlet relative to the direction of the shroud boundary layer has a strong influence on the heat transfer distribution along the shroud. In their former work, they provided another useful database in terms of isotherm maps and Nusselt number distributions for slightly different values of $C_w$ ≤ 78540 and $Re_\Omega$ ≤ $1.1\times10^6$. Poncet and Schiestel [PON07] compared favourably the predictions of the innovative RSM model of Elena and Schiestel to the experimental data of Sparrow and Goldstein [SPA76] for $C_w$=2749, 4392, 8672, $Re_\Omega$=1.56×10$^5$, 4.65×10$^5$, 9.2×10$^5$ and Prandtl numbers between 0.01 and 10. Chen et al. [CHE96] proposed a combined experimental and computational study of the heat transfer from an electrically heated rotating disk facing an adiabatic stator. A radial outflow of cooling air was used to remove heat from the disk, and local Nusselt numbers were measured using fluxmeters for flowrates up to $C_w$ = 9680 and rotational Reynolds numbers up to $Re_\Omega$ = 1.2×10$^6$. They obtained a good agreement between their measurements and the predictions of a low-Reynolds-number k-ε turbulence model. Roy et al. [ROY01] studied the convective heat transfer along the rotating disk in a rotor-stator enclosure with a mainstream flow imposed at the periphery of the cavity and a secondary air flow supplied at the centre of the stator and impinging on a small hub attached to the rotor. They conducted both experiments and RANS calculations using the RNG k-ε model within Fluent for 4.65×10$^5$ ≤ $Re_\Omega$ ≤ 8.6×10$^5$, 1504 ≤ $C_w$ ≤ 7520, $G$=0.084 and a given mainstream flowrate. Whatever the control parameters, they observed two flow regions: the source region close to the rotation axis and the core region at larger radii. In the source region, the influence of rotation on the heat transfer remains weak contrary to the secondary air flow rate, which strongly increases the heat transfer coefficient. In the core region, the convective heat transfer is mainly dominated by rotation. In that region, the authors proposed a correlation law for the local Nusselt number along the rotor: $Nu_r = 0.0195 \operatorname{Re}_r^{0.8}$.

To our knowledge, Bunker et al. [BUN90a] are the only ones to propose heat transfer measurements along both disks at the same time. They applied a thermochromic liquid crystal technique to evaluate the heat transfer along both rotating and stationary disks in a shrouded cavity for 2×10$^5$ ≤ $Re_\Omega$ ≤ 5×10$^5$, 0.025 ≤ G ≤ 0.15 and 754 ≤ $C_w$ ≤ 1626. Their results showed regions of impingement and rotational domination along the rotor with a transient region between the two characterized by low heat transfer coefficient and flow reattachement and convection regions on the stator side with an inner recirculation zone. The extent of these regions and the values of Nusselt numbers mainly depend on the aspect ratio $G$ and on the flowrate coefficient $C_w$. The heat transfer is more affected by rotation along the rotor than on the stator side. They focused also their work on the influence of the stator as an active heat transfer surface on the heat transfer distribution on the rotor. The stator heat transfer is sufficiently high to affect the local mean fluid temperature through the gap. In their second paper, Bunker *et al.* [BUN90b] investigated different geometries corresponding to parallel disk geometry or a tapered rotor with lip for different radial positions of injection. Whatever the location of injection, their results confirmed the previous ones for the dependence of the heat transfer coefficient on the control parameters $G$, $Re_\Omega$ and $C_w$. Metzger *et al.* [METZ91] used the same technique to measure the local heat transfer rate on the rotor for $G$=0.1, $Re_\Omega$ =2.71×10$^5$, 4.64×10$^5$ and 850 ≤ $C_w$ ≤ 1200. The heat transfer based on the incoming fluid temperature decreases for increasing values of the radial location up to the disk rim. Iacovides and Chew [IAC93a] examined three rotating disk cavities using three zonal modeling approaches and one model based on the mixing length hypothesis. One of three configurations corresponds to the experiments of Bunker *et al.* [BUN90a,BUN90b]. For $Re_\Omega$ =5.4×10$^5$ and $C_w$=1450, the flow is dominated by the impinging jet at the center of the cavity and for outer radii, rotational effects dominate. A large recirculation due to the inflow is observed up to $r/R$=0.7. The RANS models fail to reproduce the Nusselt number distributions, especially along the stator. The peak value observed at $r/R$=0.6 is also poorly reproduced.

Heat transfer distributions for an unshrouded rotor-stator system have been first computed by Soo *et al.* [SOO62] using an integral method. They considered an incompressible laminar flow in an unshrouded rotor-stator enclosure with or without radial outflow in two cases corresponding to a stator at either constant temperature or insulated. A constant temperature was imposed on the rotor. They proposed heat transfer distributions for relatively low values of the control parameters because of the integral method they used. Kapinos [KAP65] solved the turbulent momentum integral equations for a Batchelor-like flow with a superimposed radial outflow. Using the Reynolds analogy, he showed that the Nusselt numbers for the case with no throughflow ($C_w$=0) are the same as those for the free disk case and for an infinite flowrate coefficient ($C_w$→∞), they become independent of the rotational Reynolds number. He then compared these results with measurements of the averaged Nusselt number on a cooled rotor with a quadratic temperature profile for 0.016 < $G$ < 0.065, 5×10$^5$ ≤ $Re_\Omega$ ≤ 4×10$^6$, and a wide range of a parameter $K_0$, which combines in a complicated way $G$, $C_w$ and the radius ratio between the inner and outer cylinders. Thus, the correlation provided for the averaged Nusselt number along the rotor has only a little practical value. Kreith *et al.* [KRE63] proposed measurements for 0.12 < $G$ < 0.6, $C_w \leq 1.5 \times 10^5$, $5 \times 10^3 \leq Re_\Omega \leq 10^5$ at a given Prandtl number $Pr$=2.4. They concluded that the presence of the stator may induce a premature transition from laminar to turbulent flow on the rotor side. The radial outflow plays the same role. The flow is fully turbulent over the rotating disk for $C_w \geq 2.8 \times 10^4$. Owen *et al.* [OWE74] performed experiments where cooling air is supplied in the centre of an adiabatic stator and a parabolic temperature profile is created on the rotor by means of stationary radiant heaters. The heat transfer was calculated both by solving Laplace's equation for the rotor and by measuring temperatures of the cooling air at inlet and outlet, together with the measured windage torque of the rotor. For $0.006 \leq G \leq 0.18$, they showed that the averaged Nusselt number increases as $G$ decreases for small aspect ratios and becomes independent of $G$ for large ones. For a given value of $G$, the Nusselt number increases with increasing values of $C_w$. For large rotational Reynolds numbers, the results tend to the free-disk case, whereas for smaller values, they become independent of $Re_\Omega$. Very recently, Pellé and Harmand [PEL08, PEL09] extended their formed work on the local heat transfer on the rotor in an unshrouded rotor–stator system [PEL07] to the case of an air jet coming through the centre of the stator and impinging the rotating disk. They employed an infrared thermography technique to measure the surface temperatures on the rotor and by a thermal balance, they obtained the local convective heat transfer coefficient along the disk. Nusselt number distribution were obtained for a jet Reynolds number $Re_j$=4.16×10$^4$, a rotational Reynolds number $Re_\Omega$ between 2×10$^4$ and 5.16×10$^5$, and an aspect ratio $G$ ranging from 0.01 to 0.16. The flow along the rotating disk can be divided into zones: one dominated by the air jet near the center of the rotor and one affected by both the air jet and rotation at higher radii (Figure 24). Whatever the flow parameters, the impinging jet always enhances the heat transfer compared to the no-jet configuration, especially near the stagnation point at low rotational velocities, where the influence of the jet is preponderant. Modifying the rotational velocity has no influence on convective heat transfer because tangential velocity is very low near the center. The size of that jet-dominated area near the stagnation point also strongly depends on $G$, essentially because the width of the air jet is more significant as the spacing between the pipe outlet and the rotor gets larger. At high radii, the local Nusselt numbers were shown to depend on both jet and rotation. The influence of the aspect ratio $G$ at a given flowrate on the local heat transfer along the rotor is directly linked to the transition between a purely centrifugal flow for small interdisk spacings (*G* up to 0.02) to a Batchelor-like structure and then to a Stewartson-like structure when the aspect ratio is successively increased. In the same time, the averaged Nusselt number increases with $Re_\Omega$, whatever the value of *G*, mainly because of the local increases observed at outer radii. It decreases first as G increases up to the transition to the Batchelor regime. Pellé and Harmand [PEL09] extended their experimental database using the same apparatus (0.01≤*G*≤0.16) and measurement technique covering a wide range of jet and rotational Reynolds numbers: $Re_j$≤4.17×10$^4$ and 2×10$^4$≤$Re_\Omega$≤5.16×10$^5$. They proposed power laws to correlate the averaged Nusselt numbers along the rotor:

$$\overline{Nu_R} = 0.08 G^{-0.07} Re_j^{0.5} Re_\Omega^{0.25} \quad \text{for} \quad 0.01 \leq G \leq 0.02 \tag{62}$$

$$\overline{Nu_R} = 0.006 G^{-0.07} Re_j^{0.5} Re_\Omega^{0.5} \quad \text{for} \quad 0.04 \leq G \leq 0.08 \tag{63}$$

$$\overline{Nu_R} = 0.06 Re_j^{0.25} Re_\Omega^{0.25} \quad \text{for} \quad G = 0.16 \tag{64}$$

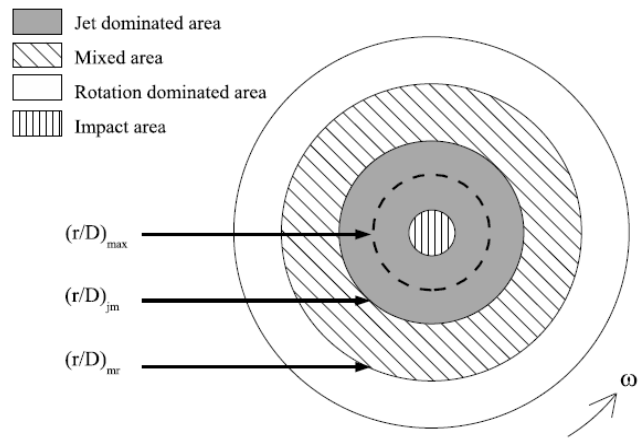

Figure 24: Four characteristic regions for the heat transfert in the case of an impinging jet onto a rotating disk in an open cavity, after Pellé & Harmand [PEL09].

The conjugate heat transfer problem including the heat transfer inside the impingement disk has been investigated quite recently by Lallave *et al.* [LAL07]. They studied the conjugate heat transfer for a confined jet impinging on a rotating and uniformly heated solid disk of finite thickness and radius. They performed extensive axisymmetric steady-state laminar calculations using a Galerkin finite element method for a wide range of control parameters including the jet and rotational Reynolds numbers, the Prandtl number, the aspect ratio of the cavity, the disk thickness to nozzle diameter ratio, and solid to fluid thermal conductivity ratio. They found that rotor materials with higher thermal conductivity maintained a more uniform temperature distribution at the solid–fluid interface. A higher Reynolds number increased the local heat transfer coefficient reducing the wall to fluid temperature difference over the entire interface. The rotational rate also increased local heat transfer coefficient under most conditions except for a disk under high spinning rate where the thermal boundary layer separates from the wall and generates an ineffective cooling. They proposed a correlation for the averaged Nusselt number along the rotor under confined jet impingement cooling of a spinning disk:

$$\overline{Nu} = 1.97619 \left(\frac{e}{D}\right)^{0.0909} \left(\frac{4Q}{\pi \upsilon D}\right)^{0.75} \left(\frac{\upsilon}{4\Omega R^2}\right)^{-0.1111} \left(\frac{\lambda_s}{\lambda_f}\right)^{-0.9} \tag{65}$$

where $\lambda_s$ and $\lambda_f$ are the thermal conductivity of the solid and the fluid respectively. In a later work, Rahman *et al.* [RAH08] extended their work on conjugate heat transfer characterization of a partially–confined liquid jet impinging on a rotating and uniformly heated disk. Their study covered also the free surface flow after exposure to the ambient gaseous medium. The dependence of the averaged Nusselt number along the impingement disk on the thermal conductivity ratio and on the jet Reynolds number is not modified compared to the case with total confinement:

$$\overline{Nu} = 1.94282 \left(\frac{e}{D}\right)^{0.111} \left(\frac{4Q}{\pi \upsilon D}\right)^{0.75} \left(\frac{\upsilon}{4\Omega R^2}\right)^{-0.0465} \left(\frac{\upsilon}{4\Omega_c R_c^2}\right)^{-0.047} \left(\frac{\lambda_s}{\lambda_f}\right)^{-0.9} \left(\frac{R_c}{R}\right)^{-0.05} \tag{66}$$

where $\Omega_c$ and $R_c$ are the rotation rate and the radius of the confinement disk. It was found that a higher flowrate increased local heat transfer coefficient reducing the interface temperature difference over the entire disk surface. The rotational rate also increased local heat transfer coefficient under most conditions.

## 2.6. Rotating flows with multi-jets

### 2.6.1 Interactions between successive jets

Depending mainly on geometrical conditions, multiple jets perform very differently from a single jet striking a target surface due mainly to two types of interactions, which do not occur in single jet systems:

- the possible jet-to-jet interaction between pairs of successive jets prior to their impingement onto the plate. This type of interference is of importance for arrays with closely spaced jets and large distances between the jets and the impingement surface.

- the interaction between the impinging jets and the flow formed by the spent fluid of the neighbouring jets. These disturbances predominantly occur for arrays with small spacing between two jets, small separation distances between the jets and the plate, and large jet velocities. The mass from one jet moves in the cross-jet flow direction, and this flow can significantly alter the performance of neighboring jets. The crossflow attempts to deflect a jet away from its impinging location on the flat plate. In situations with very strong crossflow and sufficiently large jet to plane distance, the crossflow can completely deflect the jet away from the impingement surface [KOO76].

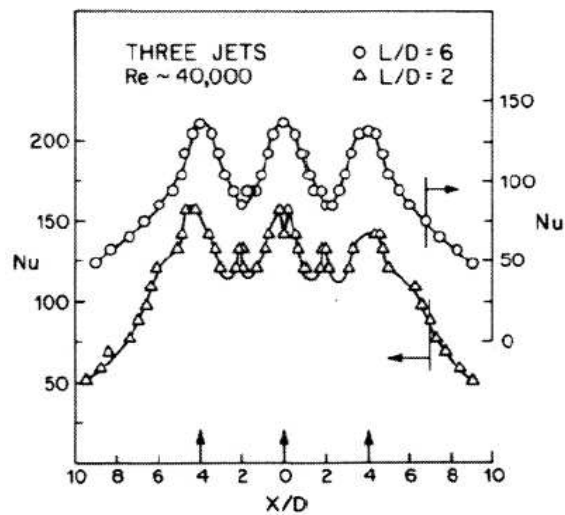

Figure 25 : Local Nusselt number of a row of three jets after [GOLD82].

Figure 25 presents the distributions of the Nusselt number for a row of three jets obtained by [GOLD82], who showed the influence of two neighbouring jets on the heat transfer coefficient for low aspect ratio e/D=2 with the appearance of secondary peaks. They disappear at larger gaps. For given separation distances, fountains as reported in Figure 26 may occur and create recirculation cells, which strongly modify the incoming jet flows. Because of all these interactions, it appears difficult to use single jet heat transfer results as base results for the design of multiple jet systems. In single jet configurations, heat transfer coefficients can be linked to the flow parameters (Reynolds, Prandtl numbers, flowrates…) using simple power-law functions. For multiple jet systems, the heat transfer rates require the consideration of a number of additional characteristic numbers.

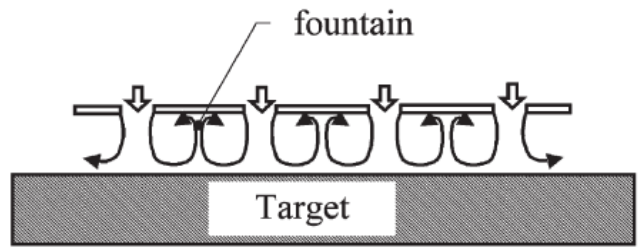

Figure 26: Typical circulation pattern in the confined jet array after [ZUC06].

We can cite, as example, the experimental investigation of Fenot *et al.* [FEN05], who used an infrared thermography technique to measure the convective heat transfer on a flat plate with either a single round jet or a row of impinging jets. They focused both on the number of jets in a row and on the influence of confinement on the heat transfer coefficient and effectiveness. Their results showed the independence of heat transfer coefficient and effectiveness from the jet injection temperature. Local Nusselt number distributions remained approximately the same as for a single jet, whatever the distance between two successive jets. A slight influence of neighbouring jets on the Nusselt number is observed midway between two jets. The distance between two successive jets affects greatly the effectiveness, which decreases more rapidly for small distances. Confinement has also only an influence on effectiveness, which decreases more rapidly without confinement.

Numerous experimental investigations have been performed up to now on heat transfer of jets impinging on a flat plate without rotation. As it is not the main object of the present review, the reader may refer to the review papers of Jambunathan *et al.* [JAM92], Viskanta [VIS93] and Weigand and Spring [WEI09] for an extensive literature surveys on jet impingement including multi-jet flows and the influence of the different parameters up to 2009. Han and Goldstein [HAN01] presented also a review of the jet impingement heat transfer in turbine systems. They have discussed the heat transfer characteristics of single jet and multiple jets, the effect of Reynolds number, Prandtl number and Nusselt number, effect of jet impingement on a curved surface and a rotating disk, and effect of angle of attack of the impinging jet and other control parameters.

### 2.6.2 Effect of rotation on the multiple jet flow problem

Although many studies have been completed for multi-jets impinging a flat surface, only few have considered the effect of rotation on impinging jet flows. As example, Epstein *et al.* [EPS85] studied the effect of rotation on impingement cooling in the leading edge of a blade. They reported that rotation decreases the impingement heat transfer. The case where the jet direction is perpendicular to rotation direction showed lower heat transfer coefficients compared to that with a staggered angle. Glezer et *al.* [GLE98] studied the effect of rotation on swirling impingement cooling in the leading edge of a blade. They found that screw-shaped swirl cooling can significantly improve the heat transfer coefficient over a smooth channel and the improvement is not significantly dependent on the temperature ratio and rotational forces. Akella and Han [AKE98] studied the effect of rotation on impingement cooling for a two-pass impingement channel configuration with smooth walls. In their case, spent jets from the trailing channel are used as cooling jets for the leading channel. They reported that whatever the direction of rotation, the heat transfer coefficient on the first pass and second pass impinging wall decreases up to 20% in the presence of rotation. Hong et *al.* [HON09] investigated the local heat and mass transfer characteristics using a naphthalene sublimation technique for various conditions of impinging jets in a rotating duct, of which an inline array of round impinging jets. The rotation changes the local heat and mass transfer characteristics due to the jet deflection and spreading phenomenon. For a ratio between the nozzle to plate spacing and the nozzle diameter equal to *e/D=6*, the jet is strongly deflected at the leading orientation, resulting in the significant decrease in heat and mass transfer. At the axial orientation, the momentum of jet core decreases slightly due to jet spreading into the radial direction and consequently, the value of stagnation peak is a little lower than that of the stationary case. The reduction of heat and mass transfers due to rotation disappears at *e/D=2*.

### 2.6.3 Multiple jets in a rotating disk system

Lee *et al.* [LEE07] performed experiments in the case of a confined rotating multi-chip module disk with round jet array impingement. They investigated the effects of many control parameters, such as the steady-state Grashof number, the ratio of jet to disk distance to nozzle diameter, the jet Reynolds number and the rotational Reynolds number on heat transfer and provided a new correlation for the stagnation Nusselt number in terms of the jet Reynolds number and normalized jet to disk distance. As compared to the single round jet case, the average heat transfer enhancement at the stagnation point of jet array impingement is 607%. For the rotating disk cases, the highest chip heat transfer occurs at the disk rim, and decreases sharply along the distance from the surface edge toward the center. To the best of our knowledge, there is at that time no relevant work dedicated to the fluid flow and heat transfer analysis of arrays of round jets impinging on a rotating disk. We can cite the early papers of Kuznetzov [KUZ64, KUZ67] and Devyatov [DEV65], who investigated the effect of multiple jet impingements on convective heat transfer for a non isothermal rotating disk in a shrouded insulated system. The single round jets were distributed along one circle either very close to the hub or at the disk rim. They considered a wide range of velocity ratios $\Omega R / W$ = 0.002 – 1 for Kuznetzov [KUZ64, KUZ67] and from 0.3 to 1.65 for Devyatov [DEV65].

This review clearly points out the lack of experimental and numerical database in the case of multiple impinging jets in rotor-stator systems. Rotor-stator disk cavities offer a simple geometry to investigate jet to jet interaction before considering more complex geometries such as blades. It suggests an interesting direction for future research.

## Conlusions

Fluid flow and convective heat transfer in rotor-stator systems are of great importance in the turbomachinery and power engineering, e.g. wind generators. A review of the most important results for the rotor-stator systems presented here elucidated the most important finding relating to investigations and predictions of convective heat transfer in predominantly outward air flow in rotor-stator cavities with and without impinging jets.

For the methodological side, the most reliable modern experimental techniques were outlined (thermocouples, infra-red cameras, laser Doppler and particle image velocimetry), and numerical were discussed, direct numerical simulation, widely used Reynolds-averaged Navier-Stokes equations techniques, large-eddy simulation approaches, self-similar solutions and integral methods.

The description of the geometries and results was structured based on the principle of mentioning first simple (free disk, disk in crossflow, single impinging jets over a stationary and a rotating disk) and finally complex rotor-stator geometries, which helps understand complex phenomena in the rotor-stator systems as a superposition of separate effects like rotation, impingement/crossflow with different angles of attack, flow regime, wall temperature profile etc.

Rotor-stator systems were outlined in full variety first without impinging jets, than with single and finally with multiple impinging jets. Maps of the flow regimes depending on the height of the gap between the rotor and stator, throughflow velocity and rotational speed help shed light onto the complex fluid flow and heat transfer phenomena in the gap. Numerous literature references to experimental investigations and numerical simulations provides a reader much space for deepening knowledge outlined in the particular parts of the review. The material is illustrated whenever possible with equations for the moment coefficient and Nusselt number.

Particular parts of the review, especially in the rotor-stator configuration sections, outline perspectives of the further extension of both the investigations and applications of the rotor-stator systems in engineering practice.